\documentclass[prb, twocolumn, floatfix, showpacs, superscriptaddress, nobibnotes]{revtex4-2}
\usepackage{graphicx}
\usepackage{color}
\usepackage{bm}
\usepackage{physics}

\newcommand{\tot}{{\rm tot}}
\newcommand{\up}{\uparrow}
\newcommand{\dn}{\downarrow}
\newcommand\figref[1]{Fig.~\ref{#1}}
\newcommand\secref[1]{Sec.~\ref{#1}}

\usepackage{amsmath}	

\begin{document}
\title{
Anomalous fractional quantization in the kagomelike Heisenberg ladder:\\
Emergence of the effective spin-1 chain
}
\author{Tomoki Yamaguchi}
\affiliation{Department of Physics, Chiba University, Chiba 263-8522, Japan}
\author{Yukinori Ohta}
\affiliation{Department of Physics, Chiba University, Chiba 263-8522, Japan}
\author{Satoshi Nishimoto}
\affiliation{Department of Physics, Technical University Dresden, 01069 Dresden, Germany}
\affiliation{Institute for Theoretical Solid State Physics, IFW Dresden, 01069 Dresden, Germany}

\date{\today}

\begin{abstract}
We study a Kagome-like spin-$1/2$ Heisenberg ladder with competing 
ferromagnetic (FM) and antiferromagnetic (AFM) exchange interactions.
Using the density-matrix renormalization group based calculations, 
we obtain the ground state phase diagram as a function of the ratio
between the FM and AFM exchange interactions. Five different phases
exist. Three of them are spin polarized phases; an FM phase and two
kinds of ferrimagnetic (FR) phases (referred to as FR1 and FR2 phases).
The spontaneous magnetization per site is $m=1/2$, $1/3$,
and $1/6$ in the FM, FR1, and FR2 phases, respectively. This can be
understood from the fact that an effective spin-1 Heisenberg chain
formed by the upper and lower leg spins has a three-step fractional
quantization of the magnetization per site as $m=1$,  $1/2$, and $0$.
In particular, an anomalous ``intermediate'' state $m=1/2$ of the
effective spin-1 chain with the reduced Hilbert space of a spin from $3$
to $2$ dimensional is highly unexpected in the context of conventional
spin-1 physics. Thus, surprisingly, the effective spin-1 chain behaves like a 
spin-1/2 chain with SU(2) symmetry. The remaining two phases are
spin-singlet phases with translational symmetry breaking in the presence
of valence bond formations. One of them is octamer-siglet phase with a
spontaneous octamerization long-range order of the system, and the other
is period-4 phase characterized by the magnetic superstructure
with a period of four structural unit cells.  In these spin-singlet phases,
we find the coexistence of valence bond structure and gapless chain. Although
this may be emerged through the order-by-disorder mechanism,
there can be few examples of such a coexistence.
\end{abstract}

\maketitle

\section{Introduction} 

The low-dimensional quantum magnets on geometrically frustrated lattices
have been studied extensively in the last decades~\cite{Diep}.
A variety of phases in such lattices originate from the macroscopically
degenerate classical ground states. The quantum fluctuations may cause
the spontaneous lift of degeneracy, which is called ``order-by-disorder''
mechanism~\cite{Villain80}. The representatives of this mechanism include
the spontaneous breaking of the lattice translational symmetry,
resulting in the valence bond solid (VBS)
state~\cite{Majumdar1969,Koga2000,Ganesh2013,Zhu2013}
as well as in the magnetic long-range
order~\cite{Jolicoeur1990,Reimers1993,Bramwell1994}, 
and that of the SU(2) spin symmetry, resulting in the ferrimagnetic (FR)
long-range order~\cite{Dmitriev2016,Yamaguchi2020}. Also, the quantum
frustrations may cause disordered quantum phases as well; the quantum
spin liquid state is one of such examples, which has been intensively
studied in the context of topological phases~\cite{Kim2000, Yamashita1246, 
Jiang2012,Han2012, Banerjee2016}.

\begin{figure}[b] 
\centering
\includegraphics[width=0.9\linewidth]{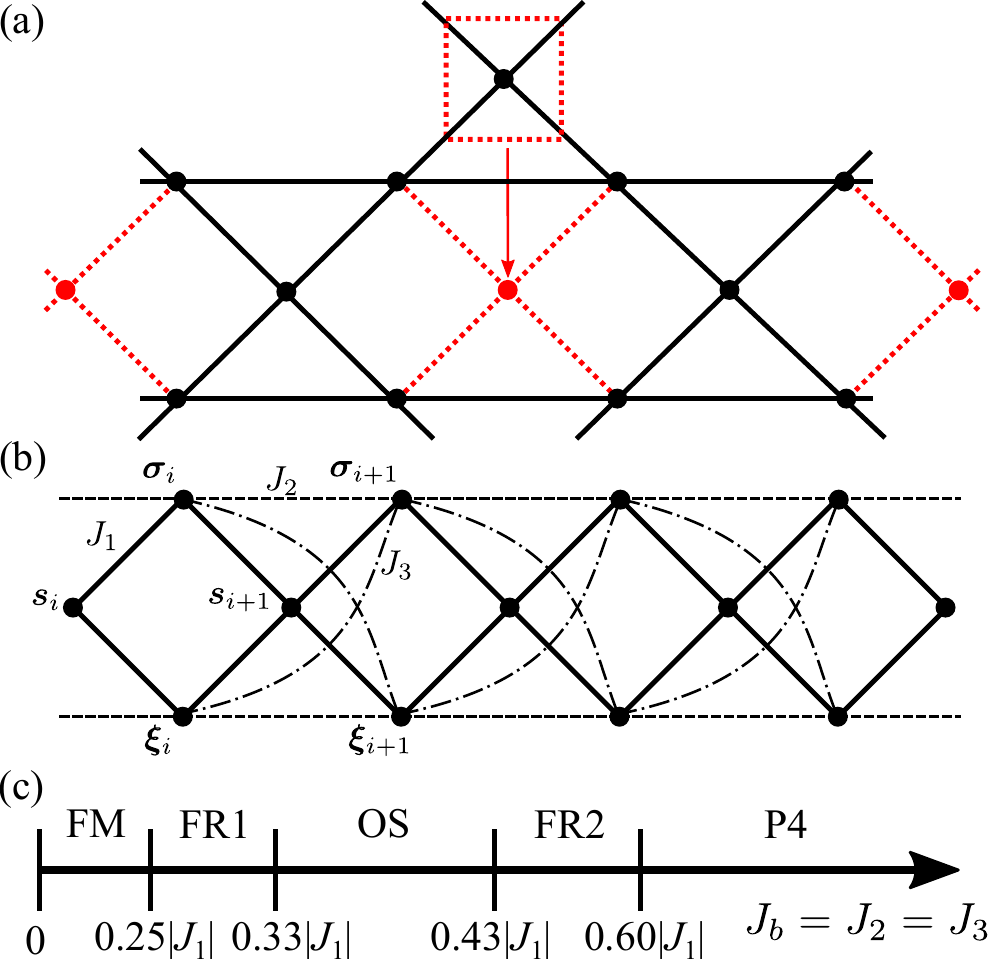}
\caption{
(a)(b) Lattice structure of the Kagome-like Heisenberg ladder with the exchange
interactions $J_1$, $J_2$, and $J_3$, where $\bm{\sigma}_i$ and $\bm{\xi}_i$
are called the upper and lower leg spins, respectively, and $\bm{s}_i$ is
called the axial (or central) spin. The Kagome-like Heisenberg ladder consists of corner-sharing triangles created by wrapping the kagome lattice to form a cylinder.
(c) Schematic ground state phase diagram of our model as a function of $J_b=J_2=J_3$ (see the main text). The abbreviations FM, FR, OS, and P4 denote
the ferromagnetic, ferrimagnetic, octamer-singlet, and period-4 phases,
respectively.
}
\label{fig:lattice}
\end{figure} 

The frustration has another interesting aspect in one-dimensional (1D) quantum
spin systems. In general, the inclusion of geometrical frustrations deserves
an inclusion of non-bipartite interactions in the system, whereby some
theorems based on the bipartite nature of the quantum spin systems
may be broken due to its quantum fluctuations. For example, FR phases have
often been discussed on the basis of the Lieb-Mattis (LM) theorem~\cite{Lieb1962},
which assumes the bipartite lattice. The LM--type FR state is considered as
the coexistence of ferromagnetic (FM) and antiferromagnetic (AFM) orders,
and the total magnetization must be integer quantized. However, in some
1D quantum spin systems, a partially-polarized FR phase, which is characterized by
a graduate change in the spontaneous magnetization, has been found~\cite{Ivanov2004, 
Hida2008, Furuya14, Yamaguchi2020}. They are kinds of non-LM--type FR phase. 
It is known that the coexistence of FM order and quasi-long-range order of 
Tomonaga-Luttinger liquid plays an essential role to realize such a partially-polarized 
FR state~\cite{Furuya14}. Thus, 1D quantum spin systems consisting frustrated 
FM and AFM interactions should provide a platform for the discovery of novel 
quantum phases of matter.

Recently, Dmitriev and Krivnov \cite{Dmitriev_J.Phys.Condens.Matter2017}
studied a spin-$1/2$ FM-AFM Kagome-like ladder using numerical
exact-diagonalization method on small clusters. The geometrical structure
of this model is shown in \figref{fig:lattice}(a), where the relevant FM-AFM
exchange pattern is assumed to be in the parameter region of $J_1<0, J_2>0$,
and $J_3>0$. When the AFM exchange interaction on the upper and lower
legs ($J_2$) and that between the two legs ($J_3$) are both small in
comparison with the nearest-neighbor FM exchange interaction ($J_1$),
this system is in a trivial FM phase. With increasing $J_2$ and $J_3$,
a phase transition from tge FM to FR states was found at
$J_2=J_3=0.25|J_1|$
based on the detailed analysis of the localized magnon states. In the FR
state, the total magnetization of the system, i.e., total spin, is $S_\tot/L=1/3$,
where $L$ is the number of sites in the system. 

The spin-1/2 Kagome-like ladder is a simplest spin model to describe
a part of magnetic properties of the half-twisted ladder 334 compounds 
Ba$_3$Cu$_3$In$_4$O$_{12}$ and 
Ba$_3$Cu$_3$Sc$_4$O$_{12}$~\cite{Koteswararao_J.Phys.Condens.Matter2012,  
Badrtdinov_Phys.Rev.B2016, Dutton_J.Phys.Condens.Matter2012, 
Kumar_J.Phys.Condens.Matter2013, Volkova_Phys.Rev.B2012}. 
These compounds exhibit similar fascinating phase diagrams with respect 
to the magnetic field and temperature. 
Of particular interest is that a series of spin-flop and spin-flip phase
transitions as a function of the magnetic field has been observed at
low temperature. To fully elucidate the nature of phase transitions,
a deeper understanding of the ground state of the spin-1/2 Kagome-like
ladder is necessary.

In this paper, we study the ground state of spin-$1/2$ FM-AFM Kagome-like
ladder using the density-matrix renormalization group (DMRG) methods.
We focus on the case of $J_2=J_3$ $(\equiv J_b)$ corresponding to the 
334 compounds~\cite{Koteswararao_J.Phys.Condens.Matter2012, 
Badrtdinov_Phys.Rev.B2016, Dutton_J.Phys.Condens.Matter2012,
Kumar_J.Phys.Condens.Matter2013, Volkova_Phys.Rev.B2012}.
Based on the numerical calculations of total spin, static spin structure factor,
spin-spin correlation functions, spin gap, dimer order parameter, and
string order parameter, we find five different phases depending on $J_b$.
Three of them are spin polarized phases with spin rotation symmetry breaking;
an FM phase and two kinds of FR phases (referred as FR1 and FR2 phases).
The spontaneous magnetization per site is $m=1/2$, $1/3$,
and $1/6$ in the FM, FR1, and FR2 phases, respectively. This can be
understood from the fact that an effective spin-1 Heisenberg chain
formed by the upper and lower leg spins has a fractional quantization
of the magnetization per site as $1$,  $1/2$, and $0$. The origin of this
anomalous quantization of the magnetization is explained in detail below.
The remaining two phases are spin-singlet phases with translational
symmetry breaking in the presence of valence bond formations.
One of them is octamer-siglet (OS) phase with a spontaneous
octamerization long-range order of the system, and the other is 
period-4 (P4) phase characterized by the magnetic superstructure
with a period of four structural unit cells. We reveal the entire ground
state phase diagram of the system as a function of $J_b/\abs{J_1}$,
as is illustrated schematically in \figref{fig:lattice}(c).

The paper is organized as follows:  In Sec.~II, we explain the spin-1/2
FM-AFM Kagome-like spin ladder and describe the numerical methods applied.
In Sec.~III, we study the total spin, spatial distribution of local magnetization,
and spin-spin correlation functions to figure out fundamental features of 
the ground state. Then, we give the detailed discussion on each phase 
in Sec.~IV. Summary is given in Sec.~V.

\section{Model and Method} 
\subsection{Model} 
The spin-1/2 Kagome-like Heisenberg ladder is defined as the three
Heisenberg chains coupled with three types of Heisenberg exchange
interactions $J_1$, $J_2$, and $J_3$. The lattice structure and geometry of
exchange interactions are illustrated in \figref{fig:lattice}(a).
The nearest-neighbor exchange interaction $J_1$ acts between the axial
(or central) and leg spins, and the interaction $J_2$ acts between the
neighboring spins in the upper and lower legs, while $J_3$ acts between
the upper and lower leg spins in diagonal positions. The Hamiltonian is 
written as
\begin{align}
  \mathcal{H} &= \sum_{i=1}^{N} \mathcal{H}_{i}, \\
  \mathcal{H}_{i} &= J_1 (\bm{s}_{i}+\bm{s}_{i+1})\cdot(\bm{\sigma}_{i}+\bm{\xi}_{i}) \nonumber\\
        &+ J_{2}\qty( \bm{\sigma}_{i}\cdot\bm{\sigma}_{i+1} + \bm{\xi}_{i}\cdot\bm{\xi}_{i+1})  \nonumber \\
        &+ J_{3}\qty( \bm{\sigma}_{i}\cdot\bm{\xi}_{i+1} + \bm{\xi}_{i}\cdot\bm{\sigma}_{i+1}), \label{eq:ham}
\end{align}
where $\bm{s}_{i}, \bm{\sigma}_{i}$, and $\bm{\xi}_{i}$ are the spin-$1/2$
operators on the axial, upper-leg, and lower-leg sites in the unit cell $i$,
respectively. The system size $L$ is given by $L=3N$, where $N$ is the
total number of the unit cells in the system.  For convenience, we also
use the notations $\bm{S}_{s, i}=\bm{s}_i$, $\bm{S}_{\sigma, i}=\bm{\sigma}_i$,
$\bm{S}_{\xi, i}=\bm{\xi}_i$, and $\bm{S}_{\alpha,i}$ $(\alpha=s,\sigma,\xi)$.
In some cases, it is beneficial to consider the system \eqref{eq:ham}
divided into the axial-spin and leg-spin parts. We refer them as
``axial-spin subsystem'' and ``leg-spin subsystem'' hereafter.

We focus on the case where $J_1$ is FM ($J_1<0$), and $J_2$ and $J_3$
are both AFM ($J_2>0$ and $J_3>0$). In particular, we further restrict
ourselves to the case at $J_2=J_3$, which corresponds to the parameters for
the half-twisted ladder 334 compounds
Ba$_3$Cu$_3$In$_4$O$_{12}$ and
Ba$_3$Cu$_3$Sc$_4$O$_{12}$~\cite{Koteswararao_J.Phys.Condens.Matter2012, 
Badrtdinov_Phys.Rev.B2016, Dutton_J.Phys.Condens.Matter2012, 
Kumar_J.Phys.Condens.Matter2013, Volkova_Phys.Rev.B2012}. 
For simplicity, we define $J_b$ as $J_b=J_2=J_3$. 
Only few theoretical studies have so far been performed on this case:
Dmitriev and Krivnov \cite{Dmitriev_J.Phys.Condens.Matter2017} discussed
the ground state manifold in the context of localized multimagnon states and
the special multimagnon complexes. They also found a discontinuous phase
transition from FM phase with $S_\tot/L=1/2$ to FR phase with $S_\tot/L=1/3$
at $J_b/\abs{J_1}=0.25$. Moreover, motivated by the experimental observations
for the half-twisted ladder 334 compounds, Kumar \textit{et al.}~\cite{Kumar_J.Phys.Condens.Matter2013} 
studied the temperature and field dependent magnetism focusing on the 
paramagnetic phase. 
Yet, little is known about the $J_b$-dependent ground-state phase
diagram of the spin-1/2 Kagome-like ladder. Therefore, it is important
to determine its ground state for better understanding of the magnetic
properties of the half-twisted ladder 334 compounds.

A key factor to understand the ground state of spin-1/2 FM-AFM Kagome-like
ladder is a formation of effective spin-1 degrees of freedom with the upper and
lower leg spins, i.e., $\bm{S}_{\sigma, i}$ and $\bm{S}_{\xi, i}$. Note that the
system [Eq.\eqref{eq:ham}] is invariant under interchange of the positions of
$\bm{S}_{\sigma, i}$ and $\bm{S}_{\xi, i}$ although it is a natural consequence
of the formation of effective spin-1 degrees of freedom. 
Therefore, the two spins $\bm{S}_{\sigma, i}$ and $\bm{S}_{\xi, i}$ are always
equivalent and the correlation between the two sites is $1/4$ independently
of $J_b/\abs{J_1}$, i.e., $\bm{S}_{\sigma, i}\cdot\bm{S}_{\xi, i}=1/4$.
As described in detail below, the combination of spin-1 degrees of freedom
and order-by-disorder mechanism yields a variety of exotic magnetisms.

Furthermore, if we consider various combinations of FM or AFM parameters
in the spin-1/2 Kagome-like Heisenberg ladder, an unexpected geometrical
frustration may give rise to a huge variety of phases. For example, even in
a simple case of $J_1>0$, $J_2>0$, $J_3=0$, several exotic phases have
been found: Waldtmann \textit{et al.}~\cite{Waldtmann2000} reported
a LM--type FR phase with magnetization per site of $m=1/6$ at
$J_2/J_1\lesssim0.5$ and a gapped period-6 VBS order at
$J_2/J_1\gtrsim1.25$. Also, M-Aghaei \textit{et al.}~\cite{M-Aghaei2018}
identified some exotic gapless phases in the region of
$0.5\lesssim J_2/J_1\lesssim 1.25$. The cases other than $J_1<0$,
$J_2>0$, $J_3>0$ are beyond the scope of this paper but the systematic
investigations appear to be intriguing future studies.

\subsection{DMRG methods} 

We employ the matrix-product-state (MPS)-based DMRG and infinite-size
DMRG (iDMRG) methods to examine the ground state of the spin-1/2
FM-AFM Kagome-like Heisenberg ladder shown in \figref{fig:lattice}(a).
We use the ITensor libraries \cite{ITensor} for numerical computations.
Due to the severely frustrating nature of the present model, the finite-size
scaling analysis may not be a straightforward task. We therefore use
three boundary conditions in complementary style: The periodic boundary
conditions (PBC) for the lattices with length up to $L=114$ (or $N=38$) are
used to avoid unphysical edge effects. The open boundary conditions (OBC)
for the lattices with length up to $L=300$ (or $N=100$) are used to
obtain accurate numerical results. And, the iDMRG method (or the
infinite-size boundary condition), which is efficient for calculating
commensurate phases, is also used to directly obtain the physical
quantities in the thermodynamic limit~\cite{McCulloch_ArXiv08042509, 
Schollwock_AnnalsofPhysics2011}. 
Since there are a number of nearly degenerate states
around the ground state due to the FM interaction $J_1$ and strong
frustration, a relatively large number of density-matrix eigenstates $\chi$
needs to be kept in the renormalization procedure to obtain accurate results.
In this paper, we keep up to $\chi=6000$ density-matrix eigenstates,
sweeping the MPSs until obtaining the ground states within the errors of
$\Delta E/L=10^{-7}|J_1|$ and $\Delta E/L=10^{-10}|J_1|$ for the DMRG
and iDMRG, respectively. Furthermore, the extrapolation is made with
respect to $\chi$ when necessary.

\section{Results of calculations} 

In this section, we study the total spin, spatial distribution of local
magnetization, and spin-spin correlation function as a preliminary step
in elucidating the ground state of our system [Eq.\eqref{eq:ham}].

\subsection{Total spin and magnetization}\label{subsec:totals} 

\begin{figure}[t] 
\centering
\includegraphics[width=0.9\linewidth]{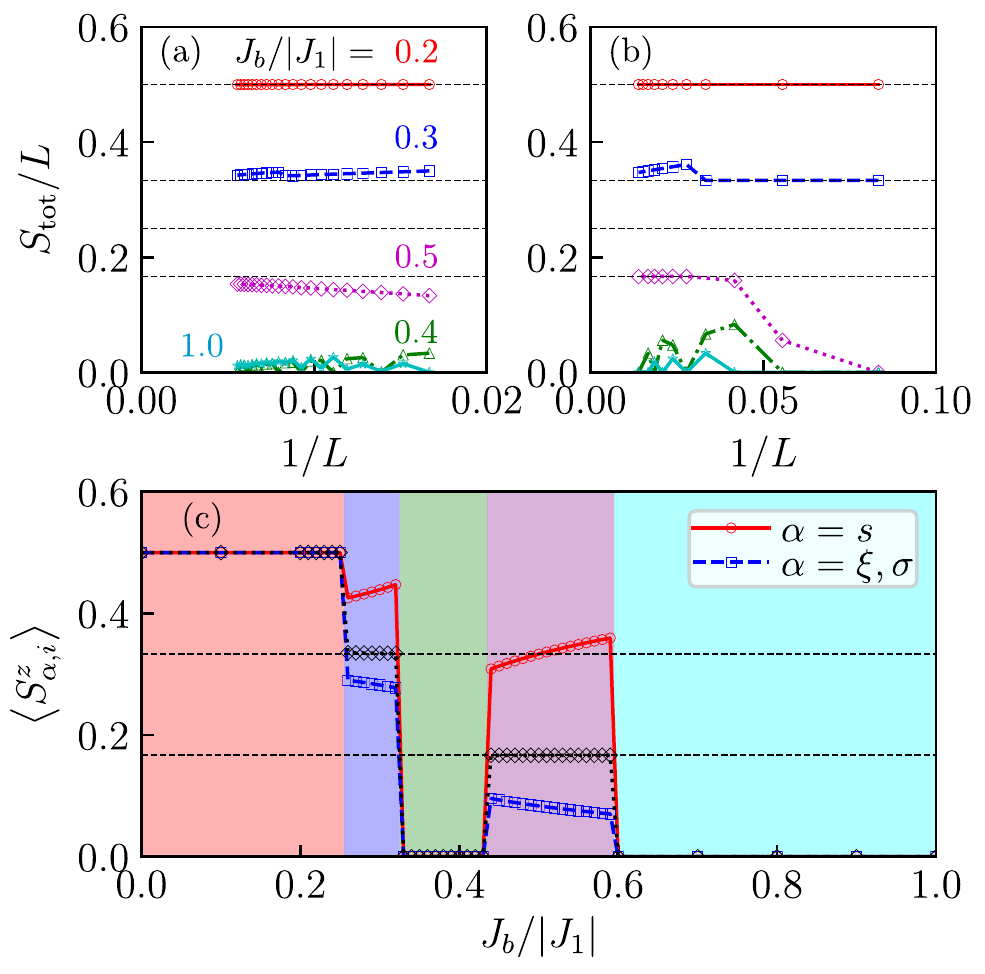}
\caption{
Finite-size scaling analysis of $S_\tot/L$ calculated using the (a) OBC and
(b) PBC, where representative parameters for each of the FM (red circles),
FR1 (blue squares), OS (green triangles), FR2 (purple diamonds), and
P4 (cyan stars) phases are chosen. (c) iDMRG results for $\ev{S_{\alpha, i}^z}$
as a function of $J_b/\abs{J_1}$; $\ev{S^z_s}$ for the axial spin and
$\ev{S^z_{\sigma}}=\ev{S^z_{\xi}}$ for the leg spins are shown by red circles
and blue squares, respectively. The $z$-component of total spin is fixed
at the value of total spin in the ground state. The averaged magnetization
per site, $m=(\ev{S^z_s}+\ev{S^z_{\sigma}}+\ev{S^z_{\xi}})/3$, is also shown
by black diamonds. Horizontal dotted lines indicate the magnetization per site
for the FR1 ($m=1/3$) and FR2 ($m=1/6$) phases calculated. The regions
of different phases are divided by different colors.
}
\label{fig:Stot}
\end{figure} 

First, in order to see the parameter dependence of spontaneous magnetization,
we calculate the total spin as a function of $J_b/|J_1|$ using the DMRG
method for OBC and PBC clusters. The total spin can be estimated
from the sum of spin-spin correlation functions over the system, namely,
\begin{align}
  \ev{\bm{S}^{2}} &= S_\tot(S_\tot+1) = \sum_{i,j}\sum_{\alpha,\beta}\ev{\bm{S}_{\alpha,i} \cdot \bm{S}_{\beta,j}},
\end{align}
where $\bm{S}$ ($=\sum_i\sum_\alpha\bm{S}_{\alpha,i}$) is the total spin operator.
In \figref{fig:Stot} (a)[(b)], the finite-size scaling analysis of $S_\tot/L$
calculated with the OBC [PBC] is performed for several values of $J_b/|J_1|$.
Although the size dependence of $S_\tot/L$ is not very straightforward
due to the strong frustration, we obtain the same extrapolated values of
$S_\tot/L$ for the OBC and PBC in the thermodynamic limit $L\to\infty$:
$S_\tot/L=1/2$, $1/3$, $0$, $1/6$, and $0$ for $J_b/|J_1|=0.2$, $0.3$, $0.4$, $0.5$, and $1.0$, respectively. This coincidence of the extrapolated values between
OBC and PBC gives a good indication to confirm the validity of the finite-size
scaling analysis. As a result, we find that the ground state of our system
is categorized into five regions by the values of $S_\tot/L$ as a function of
$J_b/|J_1|$: (i) $0 \le J_b/|J_1| \le0.25$ ($S_\tot/L=1/2$),
(ii) $0.25 \le J_b/|J_1| \lesssim0.33$ ($S_\tot/L=1/3$),
(iii) $0.33 \lesssim J_b/|J_1| \lesssim0.43$ ($S_\tot/L=0$),
(iv) $0.43 \lesssim J_b/|J_1| \lesssim0.60$ ($S_\tot/L=1/6$),
(v) $0.60 \lesssim J_b/|J_1|$ ($S_\tot/L=0$).
Since all of the obtained $S_\tot/L$ values indicate commensurate
magnetic structures, we may explicitly define the magnetic unit cell in
the whole $J_b/|J_1|$ region. It enables us to perform iDMRG simulation,
which works directly in the thermodynamic limit, by assuming proper
translational symmetry. Thus, we have confirmed the thermodynamic-limit
values of $S_\tot/L$ between three independent calculations, i.e.,
OBC, PBC, and iDMRG.

Next, we calculate the expectation value of $z$-component of spin operator
$\ev{S^z_{\alpha,i}}$ to see the real-space distribution of the local magnetization
using the iDMRG method. For this end, the $z$-component of total spin is fixed
at the value of total spin in the ground state, i.e.,
$S^z_\tot=\sum_i \sum_\alpha \ev{S^z_{\alpha,i}}=S_\tot$. In \figref{fig:Stot}
the iDMRG results for $\ev{S^z_{\alpha,i}}$ are plotted as a function of
$J_b/|J_1|$. At $J_b/\abs{J_1}<0.25$, we find
$\ev{S^z_{s,i}}=\ev{S^z_{\sigma,i}}=\ev{S^z_{\xi,i}}=1/2$ because the system
is in a trivial FM phase due to the dominant FM contribution by $J_1$.
Increasing $J_b/|J_1|$, the FM order collapses into an FR phase exhibiting
$S_\tot/L=1/3$, which is referred as FR1 phase. The phase transition between
FM and FR1 phases is of the first order. This is the same type of phase
transition as a FM-FR phase transition in the spin-1/2 FM-AFM delta
chain~\cite{Tonegawa2004, Yamaguchi2020}. In the FR1 phase, the axial spins
are nearly fully polarized and the leg spins are less polarized, i.e.,
$\ev{S^z_{s, i}}>\ev{S^z_{\sigma, i}}=\langle S^z_{\xi, i} \rangle$, which is 
consistent with the previous study~\cite{Dmitriev_J.Phys.Condens.Matter2017}.
Further increasing $J_b/\abs{J_1}$, the system goes into an unpolarized
phase with $S_\tot=0$ at $J_b/\abs{J_1}\approx0.33$ and then another type
of FR phase with $S_\tot/L=1/6$, which is referred to as FR2 phase, appears
at $J_b/\abs{J_1}\approx0.43$. That is, the narrow unpolarized phase is
sandwiched between two FR phases. This implies that the unpolarized state
would be originated from a formation of magnetic long-range order (LRO) induced 
by disorder due to the enhancement of geometrical frustration (see details in
\secref{sec:octamer}). In the FR2 phase, while the axial spins are largely
polarized, the leg spins are only weakly polarized. Although this structure seems
to be somewhat similar to that in the FR1 phase, the origins of two FR phases
are completely different as explained in \secref{sec:ferri1} and \secref{sec:ferri2}).
At larger $J_b/\abs{J_1}$ ($\gtrsim0.6$), the system exhibits an unpolarized
phase with $\ev{S^z_{s,i}}=\ev{S^z_{\sigma,i}}=\ev{S^z_{\xi,i}}=0$ again. 
This result seems natural since the AFM interaction becomes dominant.

\subsection{Spin structure factor} 

\begin{figure}[t] 
\centering
\includegraphics[width=1.0\linewidth]{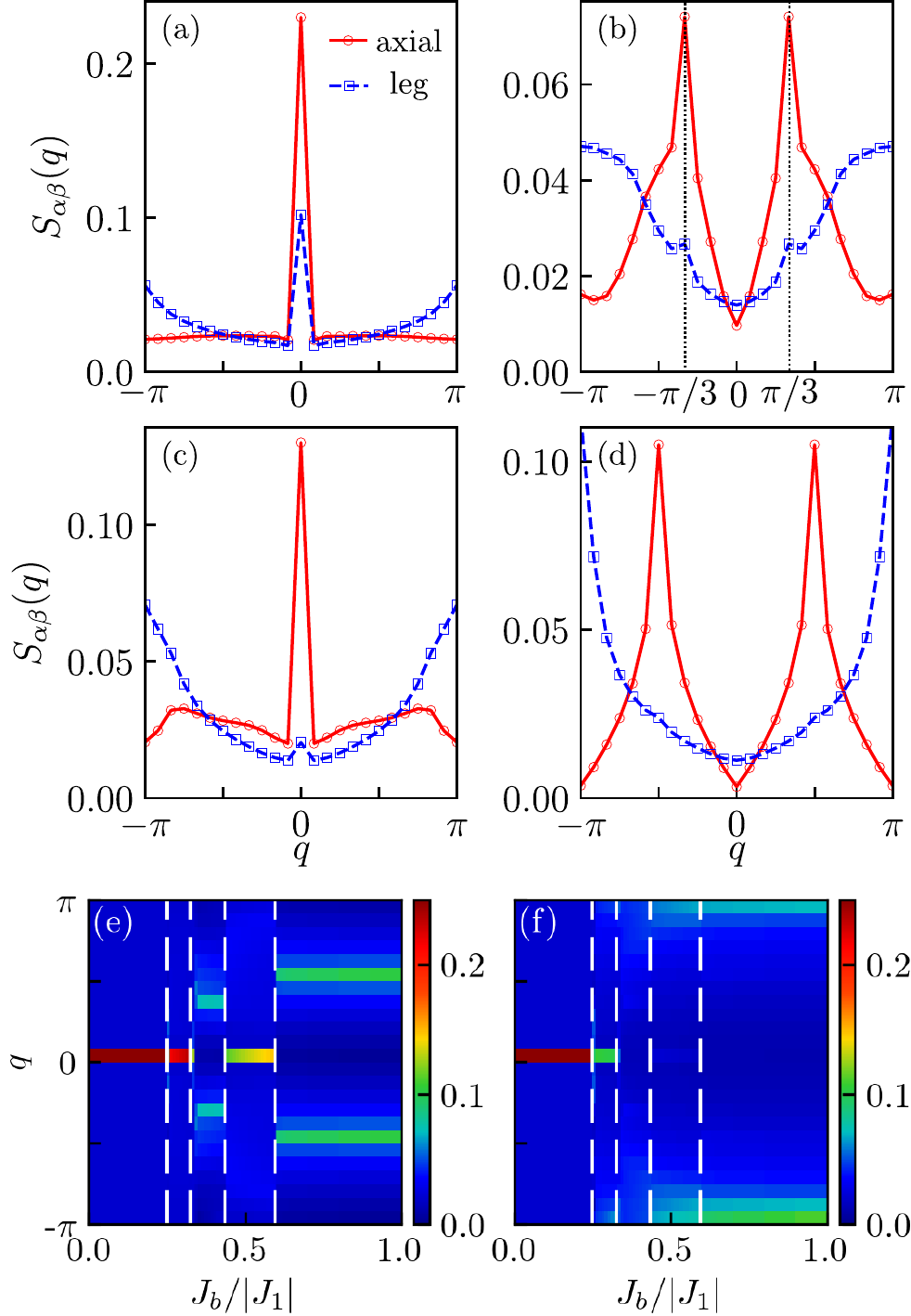}
\caption{
  DMRG results for the static spin structure factor $S_\alpha(q)$ calculated
  using $L=72$ PBC cluster. (a)-(d) $S_\alpha(q)$ with typical values of
  $J_b/\abs{J_1}$ for each phase: FR1 ($J_b/\abs{J_1}=0.3$),
  OS ($J_b/\abs{J_1}=0.4$), FR2 ($J_b/\abs{J_1}=0.5$), and
  P4 ($J_b/\abs{J_1}=1.0$). Bottom panels show the contour map of
  $S_\alpha(q)$ for (e) the axial spins $\alpha=s$ and (f) the leg spins
  $\alpha=\sigma=\xi$ as a function of $J_b/\abs{J_1}$. The four vertical
  dashed lines correspond to phase boundary in the ground state.
}
\label{fig:Sq}
\end{figure} 

To see the periodicity of magnetic structure in each phase, we calculate
the static spin structure factor which is the Fourier transform of the real-space
spin-spin correlation function
\begin{align}
  S_{\alpha\beta}(q) = \dfrac{1}{N^{2}} \sum_{i,j=1}^{N} e^{iq(r_i-r_j)}\ev{\bm{S}_{\alpha, j}\cdot\bm{S}_{\beta, i}},\label{eq:Sq}
\end{align}
where the distance between the neighboring unit cells is taken to be unity.
We here use a PBC cluster with $N=24$, i.e., $L=72$. In \figref{fig:Sq}(a)-(d), 
we show the DMRG results of $S_{\alpha\beta}(q)$ for the FR1 ($J_b/\abs{J_1}=0.3$), 
OS ($J_b/\abs{J_1}=0.4$), FR2 ($J_b/\abs{J_1}=0.5$), and P4 ($J_b/\abs{J_1}=1.0$) phases.
We note that $S_{\sigma\sigma}(q)=S_{\xi\xi}(q)=S_{\sigma\xi}(q)$ due to
the lattice symmetry.

In the FR1 phase [\figref{fig:Sq} (a)], both of $S_{ss}(q)$ and $S_{\sigma\sigma}(q)$
have a sharp $q=0$ peak reflecting the polarized axial and leg spins, and
$S_{\sigma\sigma}(q)$ has an additional relatively dull peak at $q=\pi$.
This two-peak structure of $S_{\sigma\sigma}(q)$ is a consequence of the 
anomalous value of spontaneous magnetization $S_\tot/L=1/3$
(see \secref{sec:ferri1}).

In the OS phase [\figref{fig:Sq} (b)], $S_{ss}(q)$ has its maximum value at $q=\pm \pi/3$, 
which suggests a commensurate magnetic structure with a large superlattice with 
$N=6$ unit cells, i.e., 18 sites if the LRO is assumed. The dominant fluctuations in the 
leg-spin subsystem are noted to be AFM since $S_{\sigma\sigma}(q)$ is maximum at 
$q=\pi$. 
Nevertheless, $S_{\sigma\sigma}(q)$ has also small shoulders at $q=\pm \pi/3$.
This implies that the leg spins take part in the formation of the large
superlattice structure. These are consistent with the identified OS structure
(see \secref{sec:octamer}).

In the FR2 phase [\figref{fig:Sq} (c)], the overall structures of $S_{\alpha\beta}(q)$ 
look similar to those in the FR1 phase.  The main difference is that the relative 
heights of $q=0$ and $q=\pi$ peaks are inverted. The small peak at $q=0$ corresponds 
to the weak polarization of leg spins and the larger peak at $q=\pi$ reflects the 
AFM fluctuations in the leg-spin subsystem. However, since the FM polarization 
in the axial-spin subsystem is LRO and the AFM structure in the leg-spin subsystem
is not LRO as explained in the next subsection, the $q=\pi$ peak of
$S_{\sigma\sigma}(q)$ disappears and $q=0$ peak of $S_{ss}(q)$ remains
finite in the thermodynamic limit. This peak structure is similar
to that of a FR phase in the FM-AFM delta chain~\cite{Yamaguchi2020}
because of the same origin of spontaneous magnetization (see \secref{sec:ferri2}).

In the P4 phase [\figref{fig:Sq} (d)], the peaks of $S_{ss}(q)$ and $S_{\sigma\sigma}(q)$ 
are located at $q=\pm \pi/2$ and $q=\pi$, suggesting the period of magnetic structure 
with four and two unit cells, respectively. 
Since $S_{ss}(q)$ and $S_{\sigma\sigma}(q)$ have no common peak positions,
it may be considered that the magnetic structures of axial-spin and leg-spin
subsystems are essentially separated. Therefore, although
the interaction between axial-spin and leg-spin subsystems is considered to
be weak, it is sufficient to collapse the long-range behavior of topological
string order of the leg-spin subsystem, as further discussed in \secref{sec:period4}.
In \figref{fig:Sq}(e,f) the above results are summarized as intensity plots of
$S_{ss}(q)$ and $S_{\sigma\sigma}(q)$ with $J_b/\abs{J_1}$. We find that
the dominant peak position is unchanged within each phase.  Therefore,
the periodicity would be a key factor to determine their magnetic structures.

\subsection{Spin-spin correlation function} 

\begin{figure}[t] 
\centering
\includegraphics[width=1.0\linewidth]{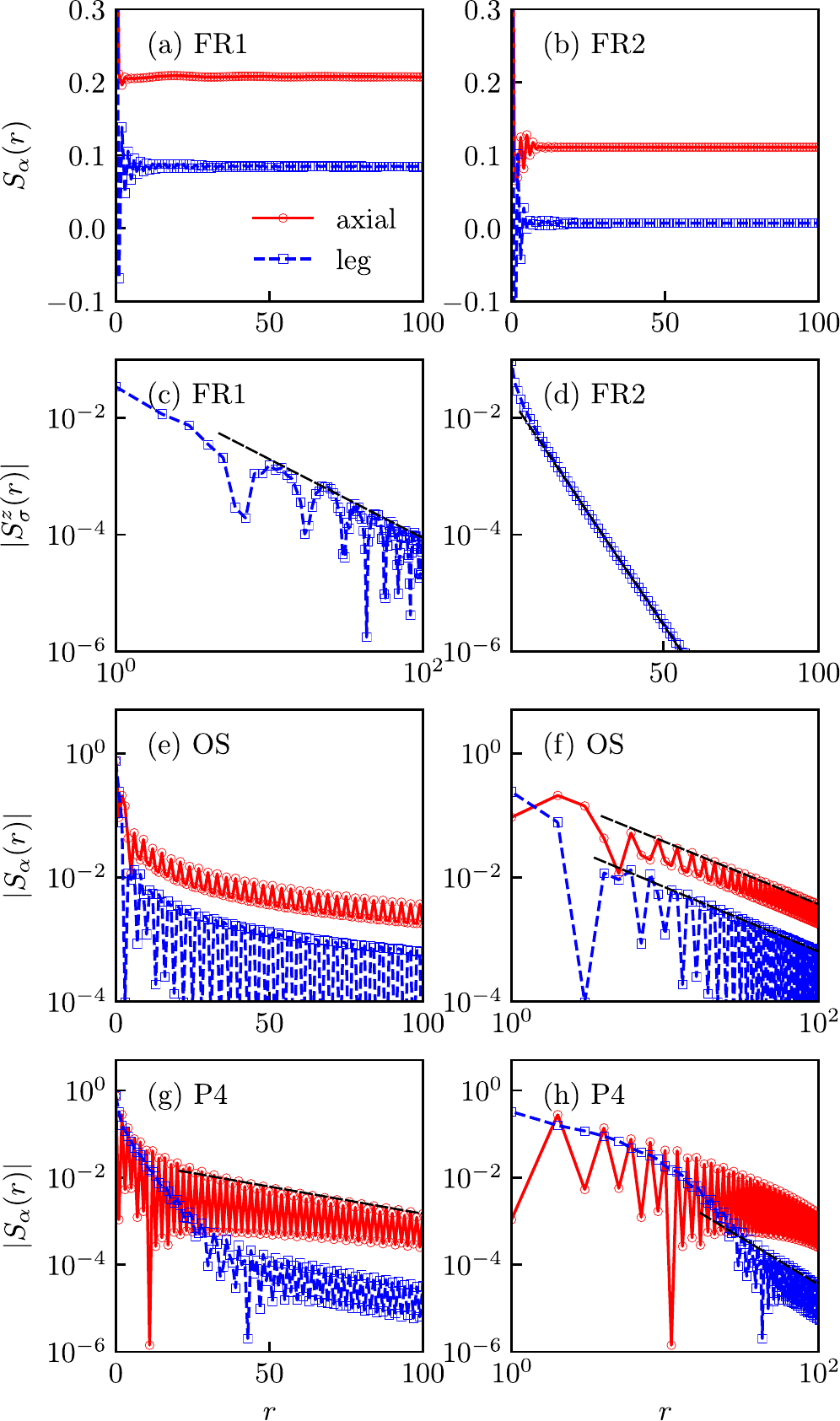}
\caption{
  iDMRG results for the spin-spin correlation function in the real space,
  where red circles and blue squares correspond to $S_s(r)$ and $S_\sigma(r)$,
  respectively. Note that $S_\xi(r)$ is equivalent to $S_\sigma(r)$.
  $J_b/\abs{J_1}$ values for each phase are chosen:
  (a,c) FR1 ($J_b/\abs{J_1}=0.3$), (b,d) FR2 ($J_b/\abs{J_1}=0.5$),
  (e,f) OS ($J_b/\abs{J_1}=0.4$), and (g,h) P4 ($J_b/\abs{J_1}=1.0$).
  In (c,d) only the oscillating part of the spin-spin correlation function is plotted.
  The dotted straight lines indicate the characteristic decays, namely, 
  a power-law decay in the log-log plot and and exponential decay in the
  semi-log plot.
}
\label{fig:Sr}
\end{figure}

In order to obtain further insight into the nature of each phase, we examine
the decay behavior of spin-spin correlation functions for each of the axial-spin
and leg-spin subsystems.The spin-spin correlation function is defined by
\begin{align}
  S_{\alpha}(r) = \ev{\bm{S}_{\alpha, i+r}\cdot\bm{S}_{\alpha, i}},
\end{align}
where the distance between neighboring unit cells is taken to be unity.
In \figref{fig:Sr}, we plot iDMRG results for the spin-spin correlation function
as a function of distance $r$. The results for $\alpha=s$ and $\alpha=\sigma$
correspond to those for the axial-spin and leg-spin subsystems, respectively.
Note that the spin-spin correlation functions for $\alpha=\sigma$ and
$\alpha=\xi$ coincide because of the lattice symmetry.

\figref{fig:Sr} (a) and (b) show $S_{\alpha}(r)$ for the FR1 and FR2
phases, respectively. We can clearly see the convergence of 
$S_\alpha(r)$ to a finite value, reflecting the FR nature of these phases
at large enough $r$. The converged value is equal to the spin polarization
squared, i.e., $S_s(r=\infty)\to\ev{S^z_{s, i}}^2$ and
$S_\sigma(r=\infty)=S_\xi(r=\infty)\to\ev{S^z_{\sigma, i}}^2$,
where $\ev{S^z_{\alpha, i}}$ is the amount of spontaneous
magnetization shown in \figref{fig:Stot}(c). We also look at the oscillating
part of spin-spin correlation function for the leg-spin subsystem,
which can be extracted via
\begin{align}
  S^z_\sigma(r)=\ev{S^z_{\sigma, i}S^z_{\sigma, i+r}}-\ev{S^z_{\sigma, i}}\ev{S^z_{\sigma, i+r}}.
\end{align}
The oscillating part $S^z_\sigma(r)$ for the FR1 and FR2 phases are
plotted in \figref{fig:Sr} (c) and (d), respectively. It is surprising that
they exhibit completely different behaviors; a power-law decay
for the FR1 phase and an exponential decay for the FR2 phase.
This means that the leg-spin subsystem is in a gapless and a gapped
ground states in the FR1 and FR2 phases, respectively. The details
are explained in \secref{sec:ferri1} and \secref{sec:ferri2}.

In \figref{fig:Sr} (e,f), the spin-spin correlation functions for the OS
phase are shown. Both $S_s(r)$ and $S_\sigma(r)$
exhibit power-law behaviors and decay approximately as
$S_{\alpha}(r) \propto 1/r$. This clearly indicates a gapless
nature of the OS state. Actually, the spinon excitation from the OS
ground state is gapless as confirmed in the next subsection.
The OS state has an LRO with alternating alignment of octamer
singlets and nearly free axial spins. Considering the period of
magnetic structure with six structural unit cells, i.e., containing 18
spins, as found by the static spin structure factor, it appears
that the nearly free axial spins are antiferromagnetically coupled
and form a spin-1/2 Heisenberg chain. Thus, the power-law behavior
with $S_{\alpha}(r) \propto 1/r$ is naively expected. Further details are
given in \secref{sec:octamer}.

In \figref{fig:Sr} (g,h), the spin-spin correlation functions for the P4
phase are shown. It is interesting that the correlation functions for the
axial-spin and leg-spin subsystems have completely different behaviors.
For the axial spins, $S_s(r)$ exhibits an exponential decay indicating
a gapped nature. In the P4 phase, all the axial spins form valence
bonds as a consequence of order-by-disorder mechanism. Thus,
the axial-spin subsystem is gapped, which is consistent with the
exponential decay of $S_s(r)$. For the leg spins, $S_\sigma(r)$ exhibits
an exponential decay at short distance and a power-law decay at long
distance. In the P4 phase, the leg-spin subsystem can be mapped
onto a spin-1 Heisenberg chain in the $S^z=0$ sector. Accordingly, 
one may expect an exponential decay of spin-spin correlation function
since the Haldane-like VBS state may be the prospective ground state.
This seems to be consistent with the exponential decay of $S_\sigma(r)$
at short distance. However, in fact, the Haldane-like VBS state is
'weakly' collapsed due to the weak interaction with the axial-spin subsystem.
Therefore, the behavior of $S_\sigma(r)$ shows a crossover from
an exponential deecay at short distance to a power-law decay at long
distance. This also means that it looks as if the Haldane-like VBS state
is stabilized at short range. In the other words, the P4 state may be
regarded as a pseudo-gapped state. Further details are given in
\secref{sec:period4} and \secref{pHaldane}.

\subsection{Spin gap} \label{sec:sgap} 
\begin{figure}[t] 
  \includegraphics[width=0.8\columnwidth]{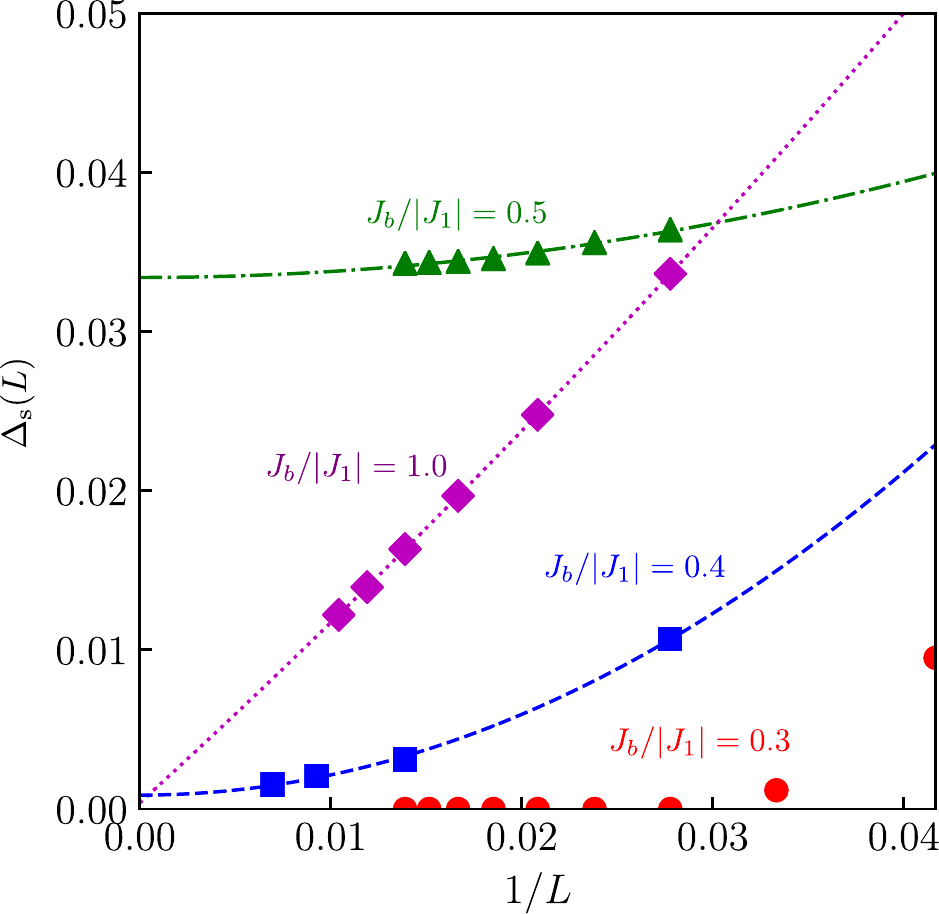}
  \caption{
    Finite-size scaling analysis of the spin gap $\Delta_\mathrm{s}(L)$.
    The representative $J_b/\abs{J_1}$ values for each phase are chosen:
    FR1 ($J_b/\abs{J_1}=0.3$), OS ($J_b/\abs{J_1}=0.4$),
    FR2 ($J_b/\abs{J_1}=0.5$), and P4 ($J_b/\abs{J_1}=1.0$).
    The lines show fitting results by polynomial functions with respect
    to $1/L$.
  }
  \label{fig:sgap}
\end{figure} 

We here examine whether a spin excitation from the ground state is gapped
or gapless in each phase. To address this issue, we define a spin gap
$\Delta_\mathrm{s}$ as the energy difference between the ground state and
the first excited state with a flipped spin:
\begin{align}
  \Delta_\mathrm{s}(L) &= E_0(L, S^z=S_{\rm tot}+1) - E_0(L, S^z=S_{\rm tot}) \\ 
  \Delta &= \lim_{L\rightarrow\infty} \Delta_\mathrm{s}(L),
\end{align}
where $E_0(L, S^z)$ is the ground state energy of the system of size $L$ and
the z-component of total spin $S^z$. For the OS and P4 phases, the ground
state is in a singlet state and we can simply take $S_{\rm tot}=0$. On the
other hand, a definite treatment is required for the FR phases since the
ground state is macroscopically degenerate. Specifically, we set
$S_{\rm tot}=L/3$ for the FR1 phase and $S_{\rm tot}=L/6$ for the FR2
phase to verify a spin excitation above their degenerate ground state.
Furthermore, to avoid underestimating the gap in association with any
free boundary effects, we adopt the PBC. In \figref{fig:sgap}, we show the size
dependence of the spin gap $\Delta_\mathrm{s}(L)$ for the
FR1 ($J_b/\abs{J_1}=0.3$), FR2 ($J_b/\abs{J_1}=0.5$),
OS ($J_b/\abs{J_1}=0.4$), and P4 ($J_b/\abs{J_1}=1.0$) phases.
The spin gap extrapolated to the thermodynamic limit is finite
only for the FR2 phase and zero for the other phases.
This can be explained as follows.

In the FR1 phase, since the axial spins are fully polarized,
the calculated spin gap corresponds to a gap in the spin excitation
spectrum of the leg-spin subsystem. As described in \secref{sec:ferri1},
the leg-spin subsystem behaves like a critical SU(2) Heisenberg chain.
Thus, the leg-spin subsystem is gapless and it is also consistent with
the power-law decay of its spin-spin correlation function $S_\sigma(r)$.

In the OS phase, our system consists of octamer singlets and residual
nearly-free spins (see \secref{sec:octamer}). The residual spins
are weakly connected and form a critical SU(2) Heisenberg chain. 
Although a finite energy is required to excite the octamer singlet,
the excitation of residual spins is gapless. A parabolic finite-size
scaling of the spin gap, i.e., $\Delta_\mathrm{s}(L)\propto1/L^2$
is a typical feature of a critical Heisenberg chain.

In the FR2 phase, the axial spins are fully polarized as in the FR1 phase.
Therefore, the finite spin gap indicates a gapful nature of the leg-spin
subsystem. It seems natural to assume that the leg-spin subsystem is
in a Haldane-like VBS state because the translational symmetry is not
broken as shown below. This is indeed a possible scenario because
the leg-spin subsystem can be mapped onto an effective spin-1 Heisenberg
chain in the $S^z=0$ sector. If this is the case, the finite spin gap
and the exponential decay of $S_\sigma(r)$ can be reasonably explained.  
A simplest way to check the presence of a Haldane-like VBS state
is to see the edge states of a open spin-1 chain~\cite{Kennedy1990}.
To achieve this, we have calculated the spin gap for two kinds of open
chains (data not shown). One is a simple open chain with remaining
upper and lower leg spins at the open edges. The other is also
an open chain but either of the upper or lower leg spins is removed at
the open edges. This is equivalent to replacing spin-1 degrees of 
freedom by spin-1/2 degrees of freedom at the edges of a spin-1 open 
chain. We have found that the spin gap is zero for the former open chain 
and is finite for the latter open chain.  
The spin gap estimated with the latter open chain coincides with that
estimated with PBC chain in the thermodynamic limit. This result
provides a solid numerical evidence for the presence of Haldane-like
VBS state.

In the P4 phase, the axial-spin and leg-spin subsystems may be 
considered separately as their spin-spin correlation functions
show completely different behaviors [see \figref{fig:Sr} (g,h)].
As described in \secref{sec:period4}, the axial-spin subsystem is
spontaneously dimerized and gapful while the leg-spin subsystem may 
be regarded as a kind of spin-1 Heisenberg chain. 
However, if the the leg-spin subsystem is in a Haldane-like VBS state
as in the FR2 phase, the spin gap $\Delta$ should be finite. Hence,
taking into account the power-law decay of $S_\sigma(r)$, the leg-spin 
subsystem seems to be in a Tomonaga-Luttinger liquid state.  
Although it appears to be true that a Haldane-like VBS state is
stabilized in the FR2 phase, the Haldane gap ($\Delta=0.034\abs{J_1}$)
at $J_b/\abs{J_1}=0.4$ is much smaller than that for the corresponding
pure spin-1 Heisenberg chain $\Delta=0.410479J_b=0.164\abs{J_1}$.
Thus, the Haldane-like VBS state in the FR2 phase seems not to be 
very stable. On the other hand, in the P4 phase, a Haldane-like VBS
state is not stabilized in a strict sense, but the tendency is seen as
an exponential decay of $S_\sigma(r)$ at short distance.  
Thus, we find it a delicate problem to identify the reason why a Haldane-like 
VBS state is stabilized in the FR2 phase but not stabilized in the P4 state, 
which is left as a future challenging issue.  

\section{Phases} \label{sec:phases}

In the previous section, based on the results of the spontaneous magnetization and
magnetic period, we have shown that the ground state of our system exhibits
five different phases, i.e., FM, FR1, OS, FR2, and P4, in the $J_b/\abs{J_1}$ space.
In this section, we determine the microscopic magnetic structure of each phase 
and reveal their origins. Especially, both the similarities and differences 
between the FR1 and FR2 states are elucidated in considerable detail.
Furthermore, we show numerical evidences for spontaneous translational symmetry
breaking, which is accompanied with valence bond formation in the OS and P4 phases.
We also clarify the reason why the OS and P4 states have a gapless spin excitation
in spite of the valence bond formation.

\subsection{Ferromagnetic (FM) phase $\bm{(0 \le J_b/|J_1| \le 1/4)}$} \label{sec:ferro}

In the limit of $J_b/\abs{J_1}=0$, our system is in a trivial FM ordered ground
state. With increasing $J_b/|J_1|$, the FM state persists up to $J_b/|J_1|=1/4$;
then, a first-order phase transition from the FM to FR1 state occurs~\cite{Dmitriev_J.Phys.Condens.Matter2017}.
The critical value of $J_b/|J_1|$ can be exactly estimated by the classical
spin wave theory. The Fourier transform of our Hamiltonian \eqref{eq:ham}
reads
\begin{align}
	H = \frac{1}{2}\sum_q J_q \bm{S}_q \cdot \bm{S}_{-q}
	\label{hamq}
\end{align}
with
\begin{align}
\nonumber
	J_q=&\frac{1}{2}[-3J_1-2J_b(1-\cos q) \\
	&\pm \sqrt{[-J_1+2J_b(1-\cos q)]^2-4J_1(1+\cos q)}],
	\label{SWT_Jq}
\end{align}
where $\bm{S}_q=(1/\sqrt{N})\sum_{\alpha,i}\exp(-iqr_i)\bm{S}_{\alpha,i}$ and
$\bm{S}_q=\bm{S}_{-q}^\ast$. Since the system contains two kinds of spins,
i.e., axial and leg spins, the dispersion is split into two branches.
The minimum position of Eq.~\eqref{SWT_Jq} changes from $q=0$ to $q=\pi$
at $J_b/|J_1|=1/4$, which is consistent with the FM critical value estimated
from the total spin. Then, Eq.~\eqref{SWT_Jq} has a minimum at $q=\pi$ in
the whole region of $J_b/\abs{J_1}>1/4$. However, as shown above,
the DMRG result for the static structure factor $S_\alpha(q)$ exhibits switching
among several peak positions depending on the value of $J_b/\abs{J_1}$.
This discrepancy implies the importance of quantum effects in this system.

\subsection{Ferrimagnetic-1 (FR1) phase $\bm{(0.25 \le J_b/|J_1| \lesssim 0.33)}$} \label{sec:ferri1}

\begin{figure}[t] 
  \includegraphics[width=0.9\columnwidth]{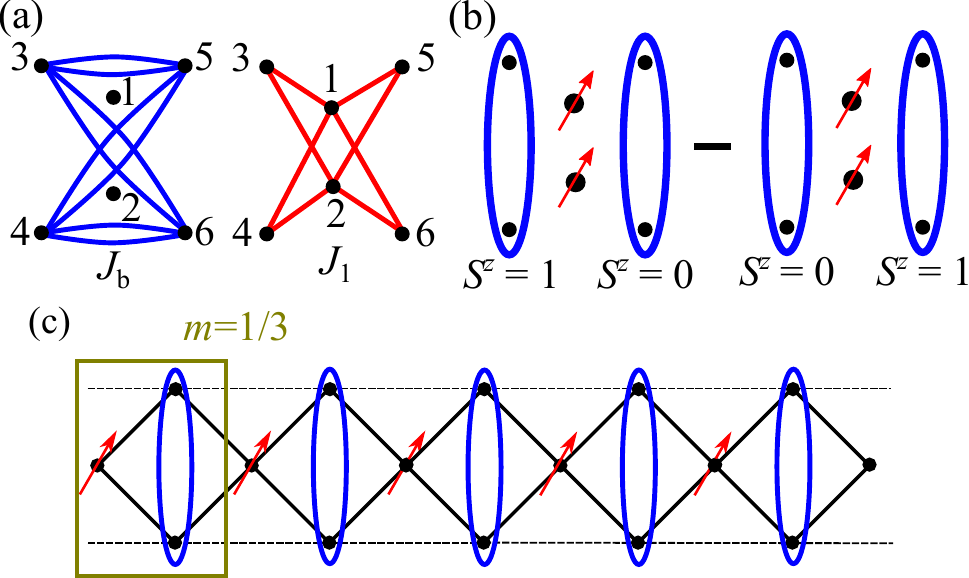}
  \caption{
    (a) A 6-site PBC cluster of the Kagome-like ladder, which may be a minimal
    unit to describe the FR1 state.
    (b) Schematic picture of the ground state of the 6-site PBC cluster
    for $0.25<J_b/\abs{J_1}<0.5$, which corresponds to the FR1 state
    with $S_\tot/L=1/3$. A blue ellipse denotes spin-triplet pair between
    an upper and a lower spin, which corresponds to an effective spin-1
    degrees of freedom and a red arrow denotes a polarized spin.
    (c) Schematic representation of the magnetic moment distributions in real
    space for the FR1 phase, where the magnetization of each spin-triplet
    pair (denoted by blue ellipse) is either $m=0$ or $m=1$. Note that the
    numbers of $m=0$ and $m=1$ are equivalent, namely, the averaged value
    is $m=1/2$, in the ground state.
  }
  \label{fig:FR1}
\end{figure} 

With increasing $J_b/\abs{J_1}$, the spontaneous magnetization $S_\tot/L$
drops down from $1/2$ to $1/3$ at $J_b/\abs{J_1}=0.25$. The value of
$S_\tot/L=1/3$, which is maintained in the region of $0.25\leq J_b/\abs{J_1}\lesssim 0.33$,
indicates a commensurate FR state. We refer to this state as FR1 state.

Let us then determine the magnetic structure of the FR1 phase.
According to the Lieb-Schultz-Mattis (LSM)
theorem~\cite{Lieb1961,Affleck1986}, a FR state with $S_\tot/L$ is allowed
only when a number
$(S-S_\tot/L)n_{\rm unit}$ is an integer, where $n_{\rm unit}$ is the number of
sites in magnetic unit cell. Since $(S-S_\tot/L)n_{\rm unit}=(1/6)n_{\rm unit}$
in the FR1 phase, $n_{\rm unit}$ needs to be a multiple of $6$. We thus consider
a 6-site periodic cluster ($L=6$) of our system \eqref{eq:ham} which may possibly
be a minimal unit to describe the FR1 state [Fig.~\ref{fig:FR1}(a)]. The Hamiltonian
of the 6-site cluster can be easily diagonalized and we find three ground
states depending on $J_b/\abs{J_1}$: $\rm(\hspace{.18em}i\hspace{.18em})$
FM state with $S_\tot/L=1/2$ ($J_b/\abs{J_1}<0.25$),
$\rm(\hspace{.08em}ii\hspace{.08em})$ FR state with
$S_\tot/L=1/3$ ($0.25<J_b/\abs{J_1}<0.5$), and
$\rm(i\hspace{-.08em}i\hspace{-.08em}i)$ singlet state with
$S_\tot/L=0$ ($J_b/\abs{J_1}>0.5$). Obviously, the state
$\rm(\hspace{.08em}ii\hspace{.08em})$ corresponds to the FR1 state.
The ground state wave function of the state $\rm(\hspace{.08em}ii\hspace{.08em})$
is exactly written as
\begin{align}
  \psi_\mathrm{GS} = \dfrac{1}{2}\ket{\up}_{1}\ket{\up}_{2}
  &\left\{\ket{\up}_3\ket{\up}_4\qty(\ket{\up}_5\ket{\dn}_6+\ket{\dn}_5\ket{\up}_6) \right. \nonumber\\
  &-\left. \qty(\ket{\up}_3\ket{\dn}_4+\ket{\dn}_3\ket{\up}_4)\ket{\up}_5\ket{\up}_6 \right\},
  \label{eq:sixsite}
\end{align}
independently of $J_b/\abs{J_1}$, where $\ket{s}_{i}$ denotes a spin state $s$
of site $i$. The site indices are assigned in Fig.~\ref{fig:FR1}(a).
For convenience, the magnetization direction is assumed to be along the z-axis,
i.e., 5 up and 1 down spins are contained in the 6-site cluster.

We here notice that two leg spins $\bm{\sigma}_{i}$ and $\bm{\xi}_{i}$, i.e.,
two spins at sites 3 and 4 (as well as 5 and 6) in the 6-site cluster, form
a spin-triplet pair, which leads to a resultant effective spin-1 site
$\bm{S}^{\rm eff}_i=\bm{S}_{\sigma, i}+\bm{S}_{\xi, i}$.
In general, a spin-triplet pair state between sites $i$ and $j$ can be mapped
onto a spin-1 degree of freedom via three $S^z$ states:
\begin{align}
  &\ket{1}_{i,j}=\ket{\up}_i\ket{\up}_j\\
  &\ket{0}_{i,j}=(\ket{\up}_i\ket{\dn}_j+\ket{\dn}_i\ket{\up}_j)/\sqrt{2}\\
  &\ket{-1}_{i,j}=\ket{\dn}_i\ket{\dn}_j
  \label{eq:spin1}
\end{align}
for $S^z=1$, $0$, and $-1$ states, respectively. Using this transformation, 
Eq.~\eqref{eq:sixsite} can be expressed as
\begin{align}
  \psi_\mathrm{GS} =
  \underbrace{\ket{\up}_{1}\ket{\up}_{2}}_{\rm (I)} \otimes  \underbrace{\dfrac{1}{\sqrt{2}}\left(\ket{1}_{3,4}\ket{0}_{5,6}-\ket{0}_{3,4}\ket{1}_{5,6}\right)}_{\rm (II)}.
  \label{eq:sixsite2}
\end{align}
Thus, the ground state wave function of state 
$\rm(\hspace{.08em}ii\hspace{.08em})$ can be separated into two parts:
(I) the fully polarized axial spins, i.e., $\ev{S^z_{s,1}}=\ev{S^z_{s,2}}=1/2$, and
(II) an antisymmetric combination of two spin states of two effective
spin-1's. It is striking that the $S^z=-1$ state $\ket{-1}_{i,j}$ is completely
projected out in part (II). Since all the leg spins $3-6$ are ferromagnetically
coupled to the polarized axial spins $1, 2$, the FM interaction $J_1$ behaves
like an external magnetic field on the leg spins, or like a uniaxial single-ion-type 
anisotropy to enhance the magnetization of each effective spin-1 
site~\cite{Botet1983}. By considering replacements
$\ket{1}_{i,j} \mapsto \ket{\up}$ and $\ket{0}_{i,j} \mapsto \ket{\dn}$,
we can easily recognize that the part (II) has the same form as a spin-singlet
pair state in spin-$\frac{1}{2}$ system. Accordingly, the two spin-1 degrees
of freedom, each of which  is exactly reduced to a SU(2) symmetry with two
states $\ket{1}_{i,j}$ and $\ket{0}_{i,j}$, are antiferromagnetically coupled.
The state of part (II) is illustrated in \figref{fig:FR1}(b).

As a result, if the system size is extended to infinity, the leg spins
$\bm{S}^{\rm eff}_i$ form a spin-1 AFM Heisenberg chain with the reduced
spin space to a SU(2) symmetry. And, the axial spins are nearly fully polarized
as also confirmed numerically in \secref{subsec:totals}. If all the axial spins
are assumed to be fully polarized, the effective spin-1 chain needs to contain
the same numbers of $S^z=1$ and $S^z=0$ sites to maintain $S_\tot/L=1/3$.
This is a natural consequence of the fact that the energy gain from
the exchange processes is maximized when the numbers of $S^z=1$ and
$S^z=0$ sites are equal,  just as the SU(2) Heisenberg chain has the lowest
energy at $S^z_\tot=0$. 
Therefore, this effective spin-1 chain behaves like
a critical SU(2) chain with the reduced Hilbert space of a spin from $3$ to $2$
dimensional. Accordingly, there is no magnetic order accompanied by
a translational symmetry breaking with alternating sites of $S^z=1$ and $S^z=0$.
This is consistent with the power-law decay of the spin-spin correlation
function for the leg-spin subsystem $S_\sigma(r)$. 

Although we assumed a magnetic unit cell with $N=6$ to fulfill the LSM
theorem, it is not the case that the FR1 state obeys the LSM theorem
because the magnetic unit cell is ill-defined in the FR1 phase.  

\subsection{Octamer-singlet (OS) phase $\bm{(0.33 \lesssim J_b/|J_1| \lesssim 0.43)}$} \label{sec:octamer} 

\begin{figure}[t] 
\centering
\includegraphics[width=1.0\linewidth]{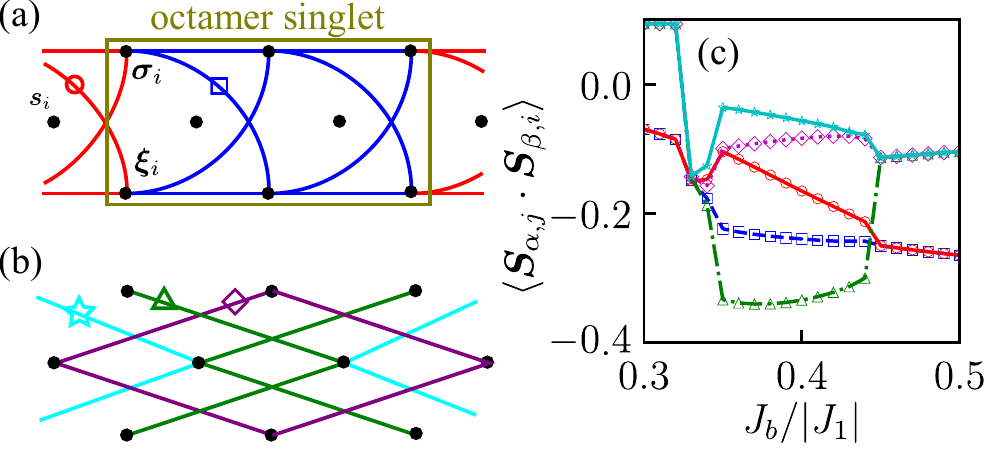}
\caption{
  (a) and (b) Structure of the octamer singlet. (c) iDMRG results for the
  spin-spin correlations as a function of $J_b/\abs{J_1}$ around the OS phase.
  The symbols and colors correspond to the lattice bonds denoted in (a) and (b).
}
\label{fig:octamer}
\end{figure} 

A narrow spin-singlet ($S_\tot=0$) phase appearing for
$0.33\lesssim J_b/\abs{J_1} \lesssim0.43$ is sandwiched between two FR phases. It means
that this $S_\tot=0$ state is not a consequence of simple melting of
FR order by AFM $J_b$ but attributed to a kind of LRO stabilization.
However, the possibility of magnetic order can be ruled out by the power-law
decay of the spin-spin correlations 
$S_\alpha(r) \propto 1/r^\alpha$ and $\ev{S^z_{s,i}}=\ev{S^z_{\sigma,i}}=\ev{S^z_{\xi,i}}=0$.
Then, the most likely LRO is a valence bond formation due to the order-by-disorder
mechanism. The representatives of such valence bond formation are
spontaneous dimerization orders in the AFM-AFM $J_1$-$J_2$ chain~\cite{Majumdar1969} and in the FM-AFM $J_1$-$J_2$
chain~\cite{Agrapidis2019}. In these systems the ground state is characterized
as a valence bond solid (VBS) state with an exponential decay of spin-spin
correlation and a finite gap in the spin excitation. And yet, in our system,
the spin-spin correlation decays in power law and the first excitation is
gapless. That is to say, although our system at $0.33\lesssim J_b\lesssim0.43$ is in an LRO
state associated with valence bond formation, it is not a simple VBS state.
In other wards, the system is not fully filled with the valence bonds.

To determine the valence bond structure, we calculate the real-space distribution
of spin-spin correlation functions $\ev{\bm{S}_{\alpha,i} \cdot \bm{S}_{\beta,j}}$.
In \figref{fig:octamer}(c) we plot the iDMRG results for 
$\ev{\bm{S}_{\alpha,i} \cdot \bm{S}_{\beta,j}}$ as a function of $J_b/\abs{J_1}$.
The symbols and colors correspond to bonds shown in \figref{fig:octamer}(a)
and (b). Although the bonds denoted by red circle and blue square are
geometrically equivalent, the correlations between these bonds are split at
$0.33\lesssim J_b/\abs{J_1} \lesssim0.43$. Similarly, the correlations among the bonds denoted
by green triangle, purple diamond, and cyan star are also split in the same
$J_b/\abs{J_1}$ region.  This result clearly indicates that the translational symmetry 
is spontaneously broken with the tripled unit cell, i.e., a 3-period magnetic unit cell
containing 9 spins. This situation may be reasonably explained by assuming
that 8 spins out of the 9 spins in the magnetic unit cell, surrounded by a
yellow box in \figref{fig:octamer} (a), form an OS state. In fact, we have
confirmed that the isolated octamer has a stable singlet ground state with
a finite excitation gap. The excitation gap has its maximum value around
$J_b/\abs{J_1}=0.4$. This means that the frustration of exchange interactions
can indeed be relieved by forming the octamer-singlet units. The details are
discussed in Appendix \ref{app:oct}.

Nevertheless, there are extra axial spins between the octamer units.
The extra axial spin thus appears in every three structural unit cells.
Considering the sharp $q=\pm\pi/3$ peaks of $S_{ss}(q)$ at 
$0.33\lesssim J_b/\abs{J_1}\lesssim 0.43$, it appears that the extra axial spins are
antiferromagnetically connected and an AFM Heisenberg chain is formed.
Since the Heisenberg chain is critical, the whole system should be gapless
in spite of gapped octamer singlets.
It is consistent with the power-law decay of the spin-spin correlations. Thus,
we conclude that the region $0.33\lesssim J_b/\abs{J_1}\lesssim0.43$ is
identified as the octamer-singlet phase. The phase transitions between
the OS and two FR phases are of the first order because these phases
have completely different symmetries.

\subsection{Ferrimagnetic-2 (FR2) phase $\bm{(0.43 \lesssim J_b/|J_1| \lesssim 0.60)}$} \label{sec:ferri2}

\begin{figure}[t] 
  \includegraphics[width=0.9\columnwidth]{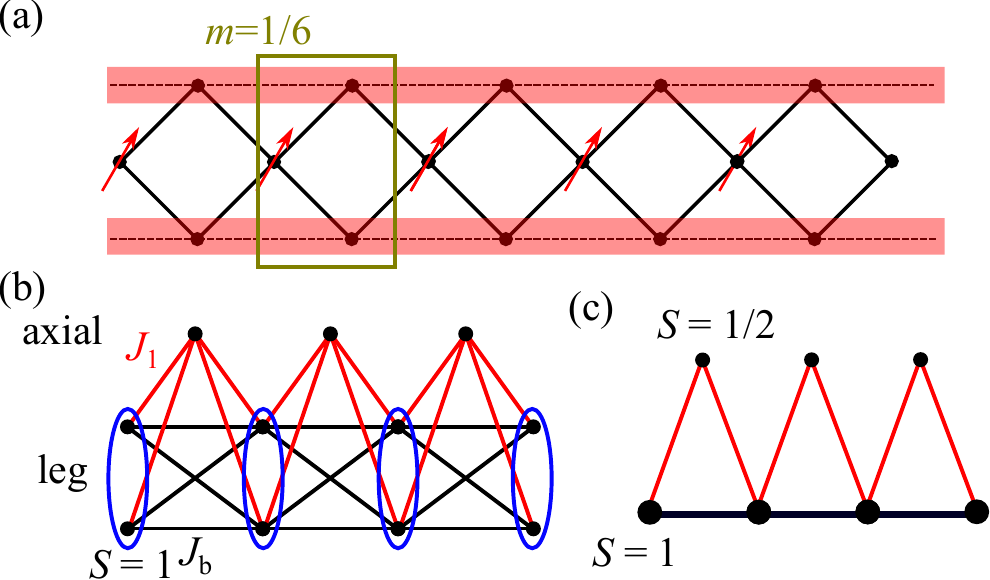}
  \caption{
    (a) Schematic representation of the magnetic moment distributions in real
    space for the FR2 phase, where the weakly polarized spins are denoted
    by red shaded area and a red arrow denotes an almost fully
    polarized spin.
    (b,c) FM-AFM delta chain as an effective model for the FR2 phase, where
    each basal spin is $S=1$ as a result of two $S=1/2$ leg spins.
  }
  \label{fig:FR2}
\end{figure} 

Another FR phase different from the FR1 phase appears at $0.43 \lesssim J_b \lesssim 0.60$,
where the total spin is $S_\tot/L=1/6$. This phase is referred to as the FR2 phase.
In the FR2 phase, the axial spins are nearly fully polarlized and the leg spins
are only weakly polarlized (see \figref{fig:Stot}). The schematic magnetic structure
of the FR2 state is sketched in \figref{fig:FR2} (a). Here, we are aware of a spin
model exhibiting a similar FR state. That is the spin-1/2 FM-AFM delta chain which
consists of nearly-fully polarlized spin-1/2 apical spins and weakly polarlized
spin-1/2 basal spins in its FR phase (we refer to this phase as ``delta-FR'' phase hereafter) 
\cite{Tonegawa2004,Krivnov2014}. Interestingly, this delta-FR order is not
associated with geometrical symmetry breaking; instead, the global spin-rotational 
symmetry is broken via the order-by-disorder
mechanism~\cite{Yamaguchi2020}.
Actually, in the FR2 phase both of the axial-axial and leg-leg spin-spin correlations
converge to finite positive values at the long distance [see \figref{fig:Sr}(b)].
This indicates a global spin-rotational symmetry breaking without any translational
symmetry breaking. Moreover, the kagome chain can be regarded as a FM-AFM
delta chain if we assume the resultant two leg spins to form $S=1$ 
[see \figref{fig:FR2} (b) and (c)]. Namely, the basal spins are not $S=1/2$ but
$S=1$. It has been confirmed that this mapping of leg chains into a single
spin-1 Heisenberg chain is exact in the limit of $J_1/J_b=0$ \cite{Kim2000}.
Nevertheless, since the FM fluctuations between the apical and basal spins
are essential to stabilize the FR state, the FR mechanism associated with
the global spin-rotational symmetry breaking should work even in the the case
of spin-1 basal chain. Thus, we may expect that the FR2 state is induced by
the same order-by-disorder mechanism as the delta-FR state.

Although the origin of the FR2 and delta-FR states are identical, there is a
discrepancy in the parameter range between them. While the delta-FR phase
maintains at the large limit of AFM interaction between the basal spins,
the FR2 phase disappears only around $J_b/\abs{J_1}=0.60$. We can think
of three possible reasons for this discrepancy: First, the $S=1$ basal
chain is more classical than the $S=1/2$ one so that the FM fluctuations
with the apical spins are more suppressed. Secondly, the weak FM order in
the effective $S=1$ basal chain may be affected by the presence of pseudo-Haldane
order (see Sec. \ref{pHaldane}). Third, even though in hindsight, the other
order-by-disorder phase, called as the period-4 phase, appears at the larger
AFM interaction $J_b/\abs{J_1}$ (see also Sec. \ref{sec:period4}).

As mentioned above, the translational symmetry is not broken in the FR2 phase.
Accordingly, the magnetic unit cell is the same as the original structural unit,
i.e., $n_{\rm unit}=3$. In that sense, one may say that the LSM condition
for a FR stabilization $(S-S_\tot/L)n_{\rm unit}=(1/3)n_{\rm unit}={\rm integer}$
is fulfilled. From this standpoint, the FR2 phase is different from the FR1 phase
which is non-LM FR one. However, the two FR phases are similar in that
the axial spins are nearly fully polarized. This fact provides us with another insight
that the fractionally quantized spin state of the effective $S=1$ chain,
consisting of the leg spins, is changed as $\ev{S}=1$, $1/2$, and $0$ in the
FM, FR1, and FR2 phases, respectively. It can be interpreted as follows: The
leg spins feel the FM correlations with the polarized axial spins like an external
magnetic field. When the AFM interaction between leg spins, i.e., $J_b/\abs{J_1}$,
is small, the leg spins have full magnetization, leading to the FM phase.
Then, the magnetization of leg spins are reduced with increasing $J_b/\abs{J_1}$.
Interestingly, the change of magnetization is not continuous but quantized.
Nevertheless, it is consistent with the Marshall-Lieb-Mattis theorem which
prohibits a ``halfway'' magnetization~\cite{Marshall1955,Lieb1962}. 
If the axial spins are directly coupled by AFM interaction,
a continuous change of magnetization may be allowed as a consequence
of the competition between the spin polarizations and a critical
Tomonaga-Luttinger–liquid behavior~\cite{Furuya14,Yamaguchi2020}.

Related to this issue, it would be informative to understand the reason
why the FR2-like state does not exist as a ground state
of the 6-site cluster discussed in Sec. \ref{sec:ferri1} [see \figref{fig:FR1}(a)].
Basically, the energy gain from the FM fluctuations between axial and leg spins
are essential to stabilize a FR state. In the FR1 state, such FM fluctuations
are simply accomplished because the system consists of axial spins with
$\ev{S}_{\rm axial}=1/2$ and leg spins with $\ev{S}_{\rm leg}=1/4$. This state
can be clearly expressed even within the 6-site cluster. On the other hand,
the leg spins need to be spontaneously polarized to stabilize the FR2 state.
To achieve this, the FM axial-leg fluctuations favoring $\ev{S}_{\rm leg}>0$ must
exceed the competing AFM intra-leg exchange fluctuations favoring
$\ev{S}_{\rm leg}=0$. However, the AFM exchange fluctuations are
always dominant with a short chain so that the FR2 state cannot exist
as a ground state of the 6-site cluster. This is one of the typical finite-size
effects and a certain chain length is required to obtain the FR2 ground state.
A similar discussion was given in the spin-1/2 delta chain \cite{Yamaguchi2020}.

\subsection{period-4 (P4) phase $\bm{(J_b/|J_1| \gtrsim 0.60)}$} \label{sec:period4} 

\begin{figure}[t] 
\centering
\includegraphics[width=1.0\linewidth]{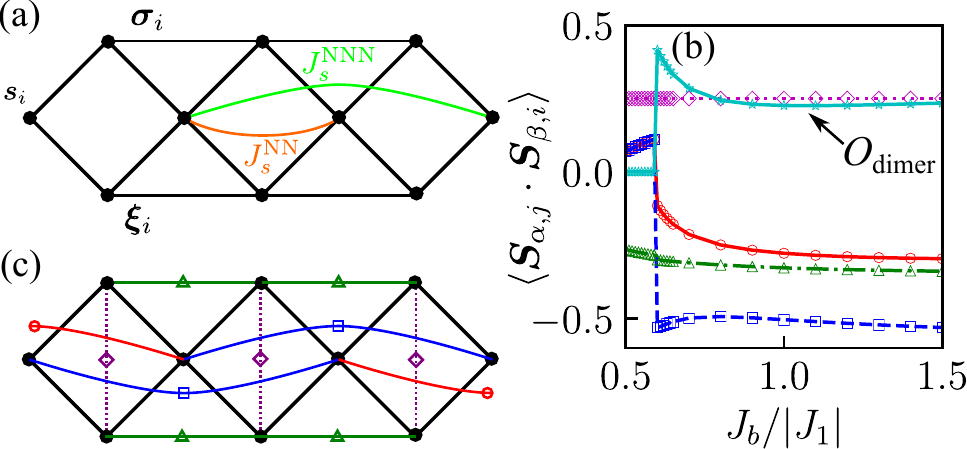}
\caption{
  (a) Structure of effective exchange couplings $J_s^{\rm NN}$ and
  $J_s^{\rm NNN}$ for the axial-spin subsystem (see text).
  (b) iDMRG results for the spin-spin correlations as a function of
  $J_b/\abs{J_1}$ in the P4 phase. The symbols and colors correspond
  to the lattice bonds denoted in (c). Cyan stars show the dimerization
  order parameter, which is estimated by Eq.~\eqref{eq:dimer} and
  is equivalent to the difference between the values denoted by
  red circles and blue squares.
}
\label{fig:period4}
\end{figure} 

At $J_b/\abs{J_1} \approx 0.6$, the system goes again into an $S_\tot=0$ phase from
the FR2 phase. As mentioned above, a fourfold magnetic structure is indicated by
the sharp peaks of $S_{\alpha \beta}(q)$ at $q=\pm\pi$ for the leg chains and
at $q=\pm\pi/2$ for the axial chain. One may naively imagine that this $S_\tot=0$
state is a spin liquid due to melting of FR2 order by AFM $J_b$. However, given
the exponential decay of spin-spin correlations between the axial spins
[\figref{fig:Sr}(e,f)], we find it not to be a simple spin liquid but a `gapped' state in
association with valence bond formation. Nevertheless, we note that the first excitation
above the ground state is in fact gapless as shown in Sec.~\ref{sec:sgap}. Although
this is seemingly puzzling, we may solve it by looking at our system divided into two 
subsystems, namely, a system of axial spins and that of leg spins.  
In fact, both of the spin structure factor $S_{\alpha \beta}(q)$ [\figref{fig:Sq}(d)]
and spin-spin correlations $S_\alpha(r)$ [\figref{fig:Sr}(e,f)] behave quite differently
between the axial-spin and leg-spin subsystems.

Let us then consider the effective models of the two subsystems. As already
mentioned above, a spin-1 AFM Heisenberg chain provides a good approximation
of the leg-spin subsystem. The maximum value of $S_{\sigma \sigma}(q)$ at
$q=\pm\pi$ [Fig.~\ref{fig:Sq}(f)] reflects the dominant AFM fluctuation. It is known
that the ground state of spin-1 AFM Heisenberg chain is a gapped state called
Haldane VBS \cite{Haldane1983} or Affleck-Kennedy-Lieb-Tasaki~\cite{Affleck1987}
state. In fact, the exponential decay of $S_\sigma(r)$ at short distance [\figref{fig:Sr}(e)]
exhibits a signature of the gapped feature. However, it turns to a power-law decay
at long distance [\figref{fig:Sr}(f)]  because the Haldane VBS state is not a perfect
long-range order due to the interaction with the axial-spin subsystem. As shown in
\figref{fig:Sr}(e) and (f), the crossover from exponential to power-law behaviors occurs
around $r_{\rm cross} \approx 35$ at $J_b/|J_1|=1$. With increasing $J_b/|J_1|$ 
the range of exponential decay, i.e., $r_{\rm cross}$, increases and the Haldane 
VBS state is recovered in the limit of $J_1/J_b=0$. This point will be further 
discussed in the next subsection.

On the other hand, a possible effective model for the axial-spin subsystem is
a frustrated FM-AFM Heisenberg chain with nearest-neighbor FM coupling
$J_s^{\rm NN}$ and next-nearest-neighbor AFM coupling $J_s^{\rm NNN}$ \cite{Bader1979}.
The ground state of this frustrated chain is either an FM state at 
$J_s^{\rm NNN}/|J_s^{\rm NN}|<1/4$ or an incommensurate spiral VBS state 
with valence bond formation between third-neighbor sites at 
$J_s^{\rm NNN}/|J_s^{\rm NN}|>1/4$~\cite{Agrapidis2019}.  
Let us estimate the effective exchange parameters within perturbation theory.
Assuming an AFM alignment of leg-spin subsystem as an unperturbed state,
the second- and third-order perturbations give
\begin{align}
  J_s^{\rm NN}\sim\frac{J_1^2}{2J_b}, ~~J_s^{\rm NNN}\sim\frac{J_1^2}{J_1+J_b}
  \label{eff_Js}
\end{align}
[see Fig.~\ref{fig:period4}(a)]. This leads to $J_s^{\rm NNN}/|J_s^{\rm NN}| \approx 2J_b/(J_1+J_b)$.
Hence, the range of $J_b/\abs{J_1}\gtrsim 0.60$ for the P4 phase corresponds to the
incommensurate spiral VBS phase of frustrated FM-AFM Heisenberg chain.
Meanwhile, as shown in \figref{fig:Sq}(d), a commensurate peak of $S_{ss}(q)$
at $q=\pi/2$ has been obtained for $J_b/\abs{J_1}=1$. In fact, the propagation
number is $q=0.4994\pi$ for $J_s^{\rm NNN}/|J_s^{\rm NN}|=1$ ($J_b/\abs{J_1}=1$)~\cite{J1J2_2012}.
Thus, the $L=72$ PBC cluster used for \figref{fig:Sq}(d) does not have high
enough resolution to detect such a tiny deviation from $q=\pi/2$. In Appendix B,
we confirm that the actual propagation number in the P4 phase is slightly less 
than $|q|=\pi/2$.

As discussed above, the ground state of axial-spin subsystem in the P4 phase
corresponds to an incommensurate spiral VBS state of frustrated FM-AFM
Heisenberg chain. We then investigate whether valence bonds are actually formed
in the P4 state. To identify the possible structure of valence bond formation,
the short-range spin-spin correlations for the bonds denoted in \figref{fig:period4}(b)
are calculated. The iDMRG results for short-range spin-spin correlations are
plotted as a function of $J_b/\abs{J_1}$ in \figref{fig:period4}(c). We find that
the spin-spin correlation between next-nearest-neighbor axial spins is uniform
in the FR2 phase ($J_b/\abs{J_1}\lesssim 0.60$), which largely splits into two 
values [denoted by red circles and blue squares in \figref{fig:period4}(c)] as soon as 
the system goes into the P4 phase ($J_b/\abs{J_1}\gtrsim0.60$). This obviously 
means that the axial-spin subsystem is spontaneously dimerized, where the valence 
bond is formed between next-nearest-neighbor axial spins coupled by 
$J_s^{\rm NNN}$ in the effective model. This is different from the fact that a 
valence bond is formed between third-neighbor sites in the frustrated FM-AFM 
Heisenberg chain.  In order to see the dimerzation strength, we define the dimerization order
parameter $O_\mathrm{dimer}$ as
\begin{align}
  O_\mathrm{dimer} &:= \abs{\ev{\bm{S}_{s, i} \cdot \bm{S}_{s, i+2}} - \ev{\bm{S}_{s, i} \cdot \bm{S}_{s, i-2}}}. \label{eq:dimer}
\end{align}
The iDMRG result for $O_\mathrm{dimer}$ is plotted as a function of $J_b/\abs{J_1}$
in \figref{fig:period4}(c). An abrupt occurrence of $O_\mathrm{dimer}$ at the boundary
between FR2 and P4 phases indicates a first-order phase transition, which is consistent
with the discontinuous changes of the total spin $S_\tot$ at the phase boundary.
With increasing $J_b/\abs{J_1}$, the dimerization order parameter decreases from
its maximum at $J_b/\abs{J_1}\approx 0.6$ and saturate at some value in the limit
of $J_b/\abs{J_1}=\infty$. This can be interpreted in terms of the frustration ratio of 
the effective model, i.e., $J_s^{\rm NNN}/|J_s^{\rm NN}|$ of the frustrated FM-AFM
Heisenberg chain. We estimate $J_s^{\rm NNN}/|J_s^{\rm NN}| = 0.75$ at $J_b/\abs{J_1}=0.6$
It is known that the frustration strength is maximum around $J_s^{\rm NNN}/|J_s^{\rm NN}| = 0.75$
in the frustrated FM-AFM Heisenberg chain~\cite{J1J2_2012}. Since the dimerization
order is a consequence of order by disorder, this result is consistent with the fact that the
dimerization order parameter is maximum around $J_b/\abs{J_1}=0.6$. Besides,
the approach of frustration ratio to $J_s^{\rm NNN}/|J_s^{\rm NN}| = 2$ at
$J_b/\abs{J_1}=\infty$ leads to the saturation of dimer order parameter.
However, the limit of $J_b/\abs{J_1}=\infty$ is not adiabatically connected to a $J_1=0$ point.
The $J_1=0$ point is singular and the dimer order no longer exists because
the axial-spin subsystem is just a group of free $1/2$ spins. This means that
the inclusion of finite $J_1$ gives non-perturbative effects in the axial-spin subsystem.
The existence of such singularity may also be expected in the AFM-AFM
Kagome-like chain~\cite{M-Aghaei2018}, which needs to be studied in the future.
In the next subsection, we will give a further discussion on this issue from the viewpoint
of the string order parameter.

As described above, our system can be considered as the direct product of gapless
leg-spin subsystem and gapped axial-spin subsystem in the P4 phase. This decoupling
of our system is further supported by the fact that the spin-spin correlation between
axial and leg spins is very small, only of the order of $0.01$. Accordingly, we can routinely
understand the completely different behaviors of spin-spin correlation functions
between the axial-spin and leg-spin subsystems, namely, the power-law decay
of $S_\sigma(r)$ [\figref{fig:Sr}(f)] and the exponential decay of $S_s(r)$ [\figref{fig:Sr}(e)]
at long distance. Note that the whole system is gapless as shown in \secref{sec:sgap}.

\subsection{Pseudo-Haldane state} \label{pHaldane} 

\begin{figure}[t] 
\centering
\includegraphics[width=0.95\linewidth]{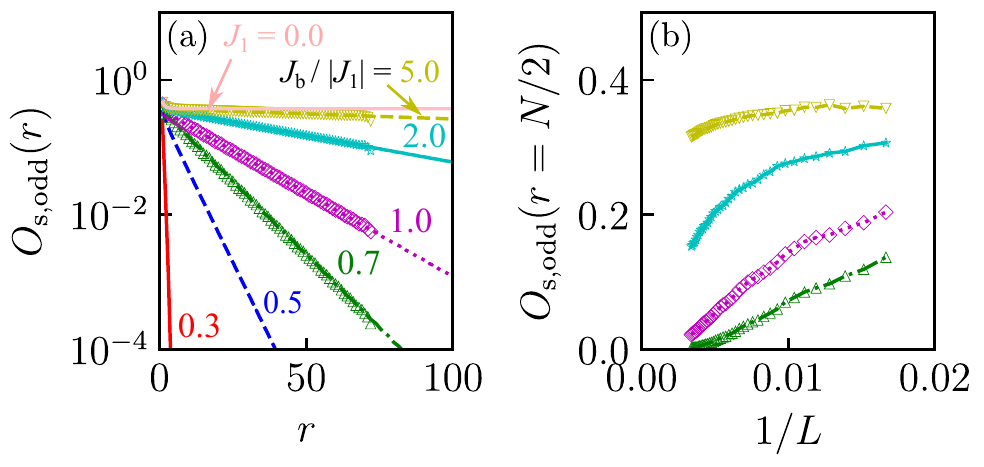}
\caption{
  (a) DMRG and iDMRG results for the z-component of a nonlocal string
  correlation function $O_{\mathrm{s, odd}}(r)$ as a function of distance $r$.
  The symbols represent OBC results with $L=300$, and lines represent
  the corresponding iDMRG results. The cyan sold line shows the result
  for $J_1=0$, where the system is decomposed into the so-called diagonal
  ladder and free axial spins (see main text). The other numbers inside the
  figure are the values of $J_b/|J_1|$. 
  (b) Finite-size scaling analysis of the string order parameter
  $O_{\mathrm{s, odd}}(r=N/2)$ calculated using OBC clusters.
}
\label{fig:SOPz}
\end{figure} 

At $J_1=0$, our system is completely decomposed into the leg-spin subsystem
and free axial spins. The leg-spin subsystem is equivalent to the so-called diagonal
ladder with uniform exchange couplings. Although the diagonal ladder has been
intensively studied in the context of columnar-dimer phase, no positive numerical
evidence for a dimerized state has so far been found in the previous
studies~\cite{Allen2000a, Starykh2004, Hung2006, Barcza2012}.
This is consistent with our results and we have further confirmed that
even the inclusion of finite FM $J_1$ does not derive any dimerized state.

Another interesting issue from the viewpoint of topological order is the hidden
Haldane VBS order~\cite{denNijs1989}. The diagonal ladder with uniform exchange
couplings, i.e, our leg-spin subsystem at $J_1=0$, is exactly mapped onto a spin-1
Heisenberg chain
\begin{align}
  \mathcal{H} = J_b \sum_{i=1}^{N} \bm{S}^{\rm eff}_i \cdot \bm{S}^{\rm eff}_{i+1}.
  \label{eq:effham}
\end{align}
Therefore, the appearance of Haldane VBS order as a hidden $Z_2 \times Z_2$
symmetry breaking is naturally expected. This order can be revealed by a string
order parameter defined through a nonlocal unitary transformation~\cite{denNijs1989,Kennedy1992-1,Kennedy1992-2,Oshikawa1992}.
Let us then consider what happens when FM $J_1$ is switched on. To study this,
we calculate the $z$-component of a nonlocal string correlation function
\begin{align}
  O_\mathrm{s, odd}(r) &= \ev{(\sigma^{z}_{i+r}+\xi^{z}_{i+r}) \prod_{i+r > k > i} e^{i\pi(\sigma^{z}_k+\xi^{z}_k)}(\sigma^{z}_{i}+\xi^{z}_{i})}.
\end{align}
The string order parameter is defined as a value of $O_\mathrm{s, odd}(r)$
at long-distance limit $r\to\infty$. In \figref{fig:SOPz} we show the iDMRG
and DMRG results for the string correlation function as a function of distance
for several values of FM $J_1$. At $J_1=0$, we can clearly confirm a long-range 
hidden VBS order with a convergence to
$O_{s, \mathrm{odd}}(r\rightarrow\infty)\simeq0.374$ at long distance,
which is consistent with the previous study~\cite{Huse1993}. Surprisingly,
just by introducing small $\abs{J_1}=0.2$ ($J_b/\abs{J_1}=5$), the correlation
function turns to an exponential decay, although its correlation length seems
to be still very large. With further increasing $\abs{J_1}$, the decay of
$O_\mathrm{s, odd}(r)$ becomes faster, keeping its exponential behavior. 
Since the $J_1=0$ point is singular for the axial-spin subsystem, it may be
a good guess that the hidden Haldane VBS order in the leg-spin subsystem
no longer exists at any finite $\abs{J_1}$. However, more precise iDMRG
analysis of $O_\mathrm{s, odd}(r)$ at semi-infinite distance, e.g., $r\sim10^5$,
is necessary to confirm it~\cite{HengSu2012}.

In \figref{fig:SOPz}(b), a finite-size scaling of $O_\mathrm{s, odd}(r=N/2)$
is shown. We recognize that the correlation length of $O_\mathrm{s, odd}(r)$
is maintained to be relatively long.at $\abs{J_1}\lesssim1$ ($J_b/\abs{J_1}\gtrsim1$).
The correlation length may roughly be estimated from the change in slope 
of $O_\mathrm{s, odd}(r=N/2)$ vs. $1/L$. For example, let us see the case 
of $J_b/\abs{J_1}=1$: The slope of $O_\mathrm{s, odd}(r=N/2)$ vs.~$1/L$ 
looks almost linear at $1/L\gtrsim0.01$, which indicates a short-range stability 
of the hidden VBS order. At $1/L\lesssim0.01$, the value of $O_\mathrm{s, odd}(r=N/2)$ 
goes toward zero with decreasing $1/L$, which suggests a collapse of the 
long-range hidden VBS order. 
From the changing point of slope, the correlation length is roughly estimated
as $r=L/3\sim33$. It is indeed reasonable that this correlation length
agrees well to the crossover distance of the spin-spin correlation functions
$S_\sigma(r)$, where the change in $S_\xi(r)$ from exponential to power-law 
behaviors occurs at $r_{\rm cross}\approx35$ [see \figref{fig:Sr}(e) and 
\secref{sec:period4}]. 
This means that it looks as if a hidden VBS order is stabilized at
$r \lesssim r_{\rm cross}$, although the long-range order is actually
collapsed by finite $\abs{J_1}$. This unique feature makes our analysis
to identify the string order parameter difficult and the use of iDMRG is
necessarily required.  A similar difficulty is seen, for example, in the study
of XY phase of spin-1 Heisenberg chain~\cite{Ueda2008}.  
This issue also reminds us of a different behavior in a similar system; 
the AFM-AFM delta chain consisting of spin-1/2 basal and spin-1 apical 
sites~\cite{Chandra2004}. In this system, the Haldane VBS phase survives 
for small apical-basal interactions. To find the reason why the different 
behaviors occur is left for the future.  

\subsection{Relevance to experiments} \label{sec:exp} 

Finally, let us comment on the relevance of our results to the experimental 
observations for Ba$_3$Cu$_3$In$_4$O$_{12}$ and Ba$_3$Cu$_3$Sc$_4$O$_{12}$. 
These materials have a similar field--temperature ($B$--$T$) phase diagram 
\cite{Kumar_J.Phys.Condens.Matter2013, Dutton_J.Phys.Condens.Matter2012, 
Volkova_Phys.Rev.B2012}: 
At $B=0$, these materials exhibit a 3D AFM order indicated
by a sharp peak of the magnetic susceptibility and a jump of the specific heat
at low $T$. It would be originated from weak AFM interchain couplings between
the Kagome-like chains.
With increasing $B$, these systems undergo a phase transition from 
the AFM to spin-flop phase, and then to a fully polarized ferromagnetic phase.
Of particular interest is that two-step phase transitions occur
in the spin-flop phase at magnetic fields of $B_1 \sim 2.0$ T and
$B_2 \sim 3.2$ T in Ba$_3$Cu$_3$In$_4$O$_{12}$~\cite{Volkova_Phys.Rev.B2012}.
At the phase transitions, the experimental magnetization is
$M/M_s \sim 0.35$ and $0.67$ for $B_1 \sim 2.0$ T and
$B_2 \sim 3.2$ T, respectively. These magnetization values
are close to those of the FR2 phase ($M/M_s=1/3$) and of
the FR1 phase ($M/M_s=2/3$). Thus, the two phase transitions
in the spin-flop phase may correspond to the instabilities 
of the two kinds of FR phase in the kagome-like chain, namely,
the difference of magnetic susceptibility between the apical
and leg spins.  Further investigation on the $B$-$T$ phase diagram 
is beyond the scope of this paper but is a fascinating open issue 
to be addressed in the future.

\section{summary} 

Using the DMRG-based techniques, we studied the spin-1/2 FM-AFM 
Kagome-like ladder with FM coupling $J_1$ between the axial and leg 
spins and  AFM coupling between the leg spins $J_b$.  Based on the 
numerical calculations of the total spin, static structure factor, spin-spin 
correlation functions, spin gap, dimer order parameter, and string order 
parameter, we found five different phases in the ground state, depending 
on $J_b/\abs{J_1}$, as summarized in \figref{fig:lattice}(c): 
$\rm(\hspace{.18em}i\hspace{.18em})$ FM ($0 \le J_b/|J_1| \le 0.25$),
$\rm(\hspace{.08em}ii\hspace{.08em})$ FR1 ($0.25 \le J_b/|J_1| \lesssim 0.33$),
$\rm(i\hspace{-.08em}i\hspace{-.08em}i)$ OS ($0.33 \lesssim J_b/|J_1| \lesssim 0.43$),
$\rm(i\hspace{-.08em}v\hspace{-.06em})$ FR2 ($0.43 \lesssim J_b/|J_1| \lesssim 0.60$),
and $\rm(\hspace{.06em}v\hspace{.06em})$ P4 ($J_b/|J_1| \gtrsim 0.60$) phases.

The FM, FR1, and FR2 phases are characterized by the spontaneous 
spin-rotational symmetry breaking. The average magnetization per site is
$m=1/2$, $1/3$, and $1/6$ in the FM, FR1, and FR2 phases, respectively.
A (nearly) fully polarization of the axial spins is a common structure
of these three phases; the difference is attributed to a fractional
quantization of the leg-spin magnetization. Since the leg-spin subsystem
can be mapped onto an effective spin-1 Heisenberg chain with spin-1
degrees of freedom formed by the upper and lower leg spins, the issue 
comes to the three-step magnetization for the effective spin-1 chain.
First, in the FM phase, the effective spin-1 chain is trivially in a full
polarization due to the dominant FM interactions. Second, in the FR1 phase,
the magnetization per site in the effective spin-1 chain is $1/2$ although
only either $0$ or $1$ is generally allowed in a stable ground state.
What happens is that an unusual critical SU(2) state with
the reduced Hilbert space of a spin from $3$ to $2$ dimensional is achieved
to maximize the energy gain from both the exchange processes and
the FM correlations with the axial spins.
Third, in the FR2 phase, the magnetization in the effective spin-1
chain should essentially be $0$ although it is actually small finite because 
of the spin-rotational symmetry breaking.  
It is surprising that, nevertheless, the Haldane-like VBS state is present
in the effective spin-1 chain. In fact, this small magnetization is
a consequence of order by disorder to stabilize the FR2 state, where
the global spin-rotational symmetry is broken, in order to lower the energy 
by the FM fluctuation between the axial-spin and the leg-spin subsystems. 
This is the same order-by-disorder mechanism as for an FR state
in the spin-1/2 FM-AFM delta chain~\cite{Yamaguchi2020}. 
Thus, an anomalous fractional quantization in a spin-1 Heisenberg chain
as an effective model for the leg-spin subsystem is a key factor to
understand the three kinds of polarized phases in the spin-1/2
Kagome-like ladder.

The remaining two phases are spin-singlet phases. One of them is
the OS phase, which is sandwiched between the FR1 and FR2 phases.
It implies that this phase is not a consequence of simple melting of
FR state but a kind of spontaneous valence bond formation from
the order-by-disorder mechanism. We performed a detailed analysis
of the short-range spin-spin correlations and identified a long-range
order with alternating alignment of octamer singlets and nearly-free
axial spins. The nearly-free axial spins are weakly antiferromagnetically
connected and a critical SU(2) chain is formed. Hence, this state is
gapless and also consistent with the period of magnetic structure
with six unit cells, i.e., containing 18 spins, indicated by sharp peaks
at $q=\pm\pi/3$ in the static spin structure factor for the axial-spin
subsystem.  We should note that this AFM correlation is not LRO.  
The other spin-singlet phase is the P4 phase characterized
by the magnetic superstructure with a period of four structural unit cells. 
This phase appearing at large AFM coupling $J_b/\abs{J_1}$ may
be accounted for by the melting of FR state. Still, the magnetic
structure is not very simple. The axial-spin subsystem is gapped with
spontaneous dimerization. The leg-spin subsystem behaves like
a spin-1 Heisenberg chain at short distance and like a critical chain
at long distance. Accordingly, the string correlation function exhibits
an exponential decay but with a long correlation length. The string
order is recovered in the limit of large $J_b/\abs{J_1}$. Thus,
in the OS and P4 phases, we detected the coexistence of valence
bond structure and gapless chain. Although this may be emerged
through the order-by-disorder mechanism, there can be few examples
of such a coexistence.

We also discussed the relevance of the spin-1/2 FM-AFM Kagome
ladder to the experimental observations of a series of field-induced
spin-flop transitions in Ba$_3$Cu$_3$In$_4$O$_{12}$ and
Ba$_3$Cu$_3$Sc$_4$O$_{12}$. Since the magnetization values at
the two spin-flop transitions are close to those of the FR2 phase 
($M/M_s=1/3$) and of the FR1 phase ($M/M_s=2/3$). Thus, the two
phase transitions in the spin-flop phase may correspond to the
instabilities of the two kinds of FR phase. Further investigations 
of the spin-1/2 Kagome-like ladder on the $B$-$T$ parameter space 
are highly desired in the future.

\section*{ACKNOWLEDGEMENTS}
We thank Ulrike Nitzsche for technical support. 
This work was supported by Grants-in-Aid for Scientific Research 
from JSPS (Projects Nos.~JP17K05530, JP19J10805, and JP20H01849) 
and by SFB 1143 of the Deutsche Forschungsgemeinschaft 
(Project-id 247310070).  T.Y.~acknowledges financial support from the 
JSPS Research Fellowship for Young Scientists.  
Some part of numerical computations were carried out on XC40 
at YITP in Kyoto University.  

\appendix

\begin{figure}[tbh]
  \includegraphics[width=0.9\columnwidth]{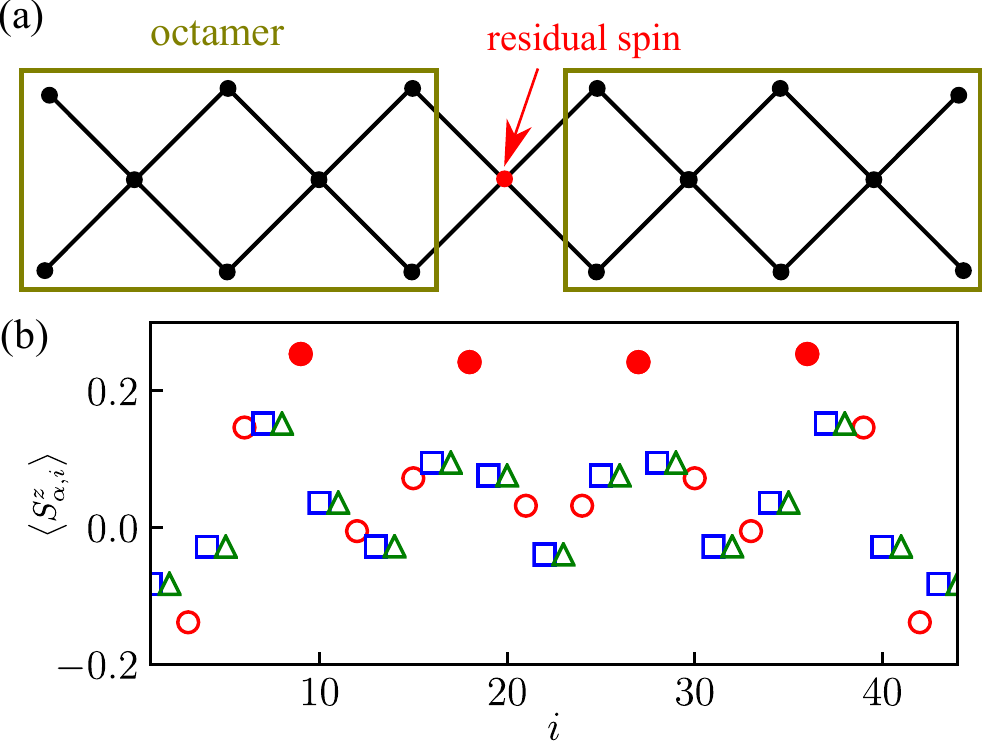}
  \caption{
    (a) A possible realization of octamer-singlet state with an open cluster.
    (b) DMRG results for the expectation value of the $z$-component of spin
    operator $\ev{S_{i}^z}$ in the $S^z=2$ sector, as a function of site index $i$.
    $J_b/\abs{J_1}=0.4$ and $L=44$ are chosen.
    Red circles, blue squares, and green triangles represent $s$, $\sigma$,
    and $\xi$ site, respectively. Filled red circles denote the values of
    $\ev{S_{i}^z}$ for nearly free spins at $i=9n$, where $n$ is an integer.
  }
  \label{fig:moment}
\end{figure}

\section{Spin polarization in the octamerized phase}\label{app:oct}

In the main text, we have argued that our system \eqref{eq:ham} is
in the octamer-singlet ground state at $0.33\lesssim J_b/\abs{J_1} \lesssim 0.43$. 
The stabilization of this state is attributed to a spontaneous
octamerization of the system with translational symmetry breaking.
More specifically, as sown in \figref{fig:moment}(a) an octamer singlet
and a nearly free spin are aligned alternately. Here, in order to
further confirm the octamerization, we give another numerical
evidence for the formation of octamer singlets with the use of
the difference of magnetic susceptibility between the octamer
singlet and nearly free spin. We apply the open boundary conditions
and keep the total number of site $L$ equal to $L=9n-1$, where
$n$ is an integer value, to be consistent with the octamer-singlet
ground state. With this setup, there are $n-1$ nearly free spins.
If one spin is flipped in the octamer-singlet ground state, the nearly
free spin out of 9 spins should be firstly polarized. 

In \figref{fig:moment}(b), we show an expectation value of the
z-component of spin operator $\ev{S^z_i}$ at $J_b/\abs{J_1}<0.4$,
where the system size is set to be $L=44=9\times5-1$ with
$S^z_\tot=2$. As we intuitively expected, a substantial polarization
is seen at four sites corresponding to the nearly free spins
(red filled circles); the other spins forming the octamer singlets
are much less polarized  (open symbols). This result supports
the octamerization of our system in the intermediate AFM coupling
regime $0.33\lesssim J_b/\abs{J_1}\lesssim 0.43$. 

We note that the nearly free spins are effectively antiferromagnetically 
coupled with each other since the period of magnetic
modulation is twelve sites because of sharp peaks of the apical-spin
structure factor $S_{ss}(q)$ at $q=\pm \pi/3$. Nevertheless, the
AFM chain consisting of nearly free spins is critical, which is consistent
with the gapless behaviors shown in the main text.

\begin{figure}[tbh]
  \includegraphics[width=0.9\columnwidth]{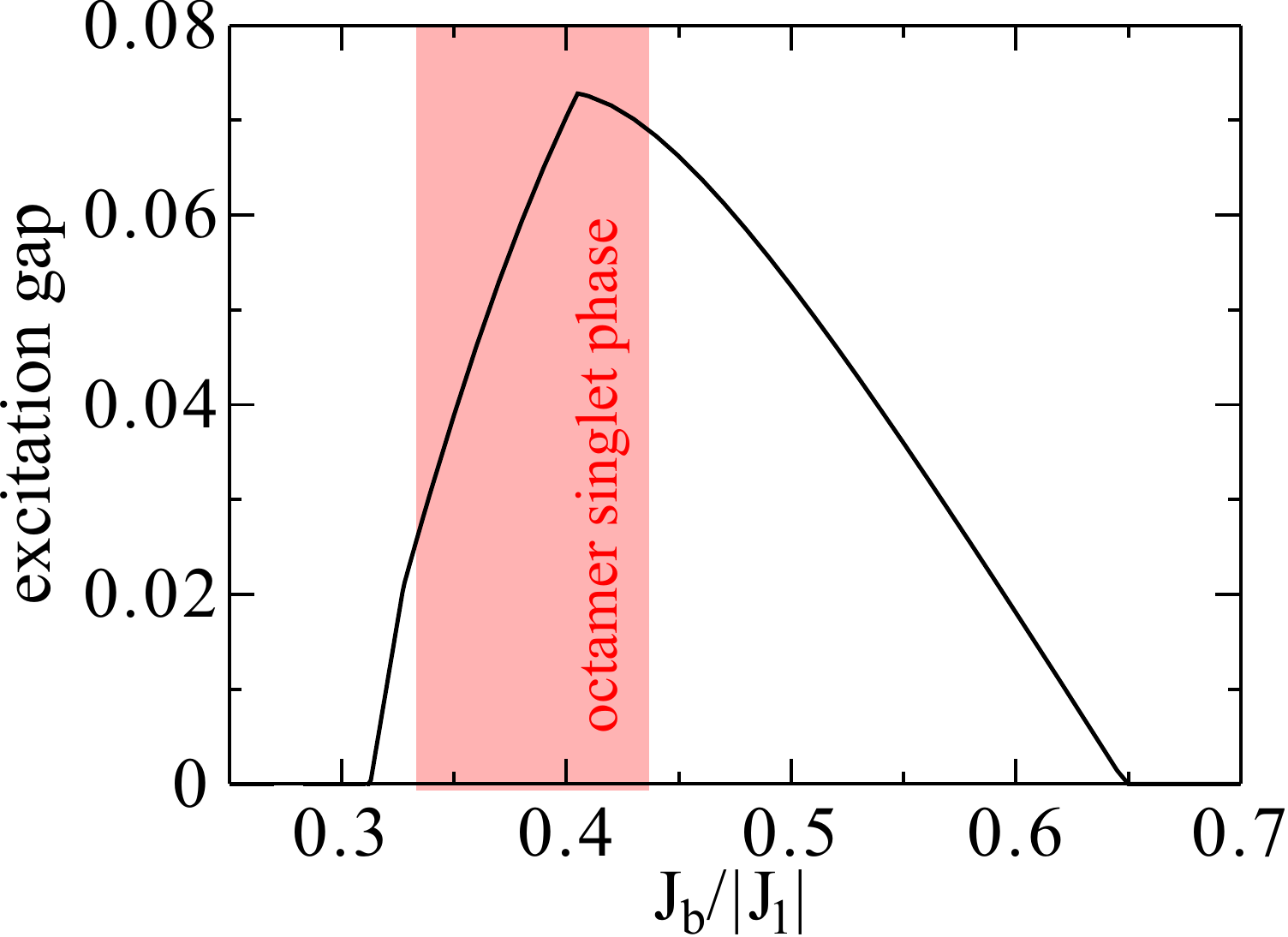}
  \caption{
Excitation gap, defined as an energy difference between the $S^z=0$ 
ground state and $S^z=1$ excited state, of isolated octamer. The red
shaded area represents the octamer-singlet phase of the kagome-like
chain.
  }
  \label{fig:os_gap}
\end{figure}

\section{Excitation gap of the isolated octamer singlet}\label{app:octgap}

In the main text, we have claimed a spontaneous formation of
octamer singlets through the order-by-disorder mechanism
at $0.33\lesssim J_b/\abs{J_1}\lesssim 0.43$. However, in fact, 
it is a nontrivial question whether each octamer has a gapped excitation
from its singlet ground state. To confirm this, we calculate the
lowest energy of isolated octamer in the $S^z=0$ and $S^z=1$
sectors. We then find that the energy for $S^z=0$ sector is
indeed lower than that for the $S^z=1$ sector at
$0.31 \lesssim J_b/\abs{J_1} \lesssim 0.65$. This indicates that
the isolated octamer has a singlet ground state in this $J_b/\abs{J_1}$
range. In \figref{fig:os_gap}, the energy difference between
the $S^z=0$ ground state and $S^z=1$ excited state is plotted 
as a function of $J_b/\abs{J_1}$, which corresponds to an excitation 
gap of the isolated octamer.  The excitation gap has a maximum around 
$J_b/\abs{J_1} \sim 0.4$. In the octamer-singlet phase of the kagome-like 
chain, the order-by-disorder mechanism works to lower the frustration 
energy by spontaneously forming octamer singlets. Thus, it appears 
that the excitation energy directly reflects the stability of octamer-singlet 
phase.  It is quite reasonable that the octamer-singlet phase exists 
in a $J_b/\abs{J_1}$ region, where the excitation gap of isolated octamer 
is maximized.

\begin{figure}[t]
  \includegraphics[width=0.9\columnwidth]{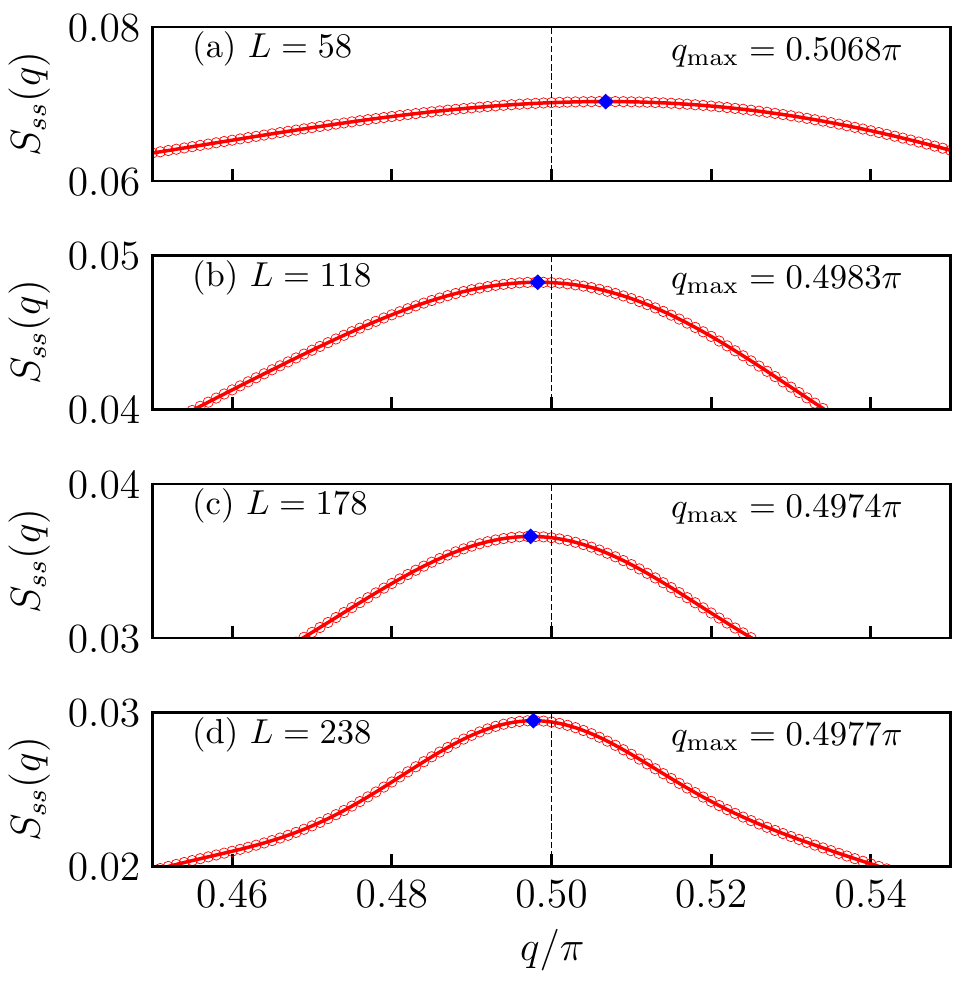}
  \caption{
    DMRG results of static structure factor $S_{ss}(q)$ for the axial spins
    in the kagome-like chain at $J_b/\abs{J_1}=0.8$, where the open
    boundary conditions are used. The maximum position is marked by
    blue diamond and the corresponding $q$-value is shown as $q_{\rm max}$.
  }
  \label{fig:Sqopen}
\end{figure}

\section{Nearly commensurate spin structure in the period-4 phase}\label{app:incomm}

The period-4 state may be characterized by a commensurate peak
of $S_{ss}(q)$ at $q=\pm \pi/2$. This means that the magnetic period 
is four structural unit cells, containing axial spins. On the other hand,
as described in the main text, the apical-spin subsystem can be
effectively mapped onto the so-called ferromagnetic-antiferromagnetic
$\tilde{J}_1$-$\tilde{J}_2$ chain in the period-4 phase at $J_b/\abs{J_1}\gtrsim 0.6$. 
In the main text, $\tilde{J}_1$ and $\tilde{J}_2$ are referred to as
$J_s^{\rm NN}$ and $J_s^{\rm NNN}$, respectively. From the
perturbative analysis [Eq.~\eqref{eff_Js}], the range of frustration
ratio is estimated to be $\tilde{J}_2/\abs{\tilde{J}_1}=0.75-2$ for
$J_b/\abs{J_1}\gtrsim 0.6$. Although the propagation number is known
to be very close to $q=\pm \pi/2$ in this $J_2/\abs{J_1}$ range,
a true commensurate modulation with $q=\pm \pi/2$ is achieved
only in the limit of $J_2/\abs{J_1}=\infty$~\cite{Bursill1995}.
If this is the case, the precise peak position of $S_{ss}(q)$ in the
period-4 phase could be slightly shifted from $q=\pm \pi/2$.
To check this, we calculate the static spin structure factor $S_{ss}(q)$ 
[Eq.~\eqref{eq:Sq}] with open chains. The use of open boundary conditions 
enables us to tune the propagation number continuously as a function 
of $J_b/\abs{J_1}$ in the situation of ill-defined momenta.  
In \figref{fig:Sqopen}, the DMRG results of static structure factor
$S_{ss}(q)$ for the axial spins in the kagome-like chain at
$J_b/\abs{J_1}=0.8$ are shown. Note that the parameter
$J_b/\abs{J_1}=0.8$ corresponds to $\tilde{J}_2/\abs{\tilde{J}_1}=0.88$
in the effective $\tilde{J}_1$-$\tilde{J}_2$ chain. We can see
that the peak position is obviously shifted from $q=\pi/2$. Although
we have not performed the finite-size scaling analysis, the propagation
number seems to converge to $q_{\rm max}=0.497\pi-0.498\pi$
in the thermodynamic limit $L\to\infty$. It is also convincing that
this $q_{\rm max}$ value agrees very well with propagation number
$q=0.498\pi$ for $\tilde{J}_2/\abs{\tilde{J}_1}=0.88$ in the
$\tilde{J}_1$-$\tilde{J}_2$ chain~\cite{J1J2_2012}.

\bibliography{kagome}

\end{document}